%% file: main.tex
\documentclass[conference]{IEEEtran}
\usepackage{cite}
\usepackage{amsmath,amssymb,amsfonts}
\usepackage{algorithmic}
\usepackage{graphicx}
\usepackage{textcomp}
\usepackage[table]{xcolor}
\usepackage{fancyhdr}
\usepackage[hyphens]{url}

\usepackage{authblk}

\usepackage[final]{microtype}
\usepackage{flushend}
\usepackage{booktabs}
\usepackage{array}
\usepackage[subrefformat=parens,labelformat=parens]{subfig}   
\usepackage{tablefootnote}
\usepackage{tabularx}
\usepackage{diagbox}

\newcommand{\todo}[1]{}
\renewcommand{\todo}[1]{{\color{red} TODO: {#1}\\}}
\usepackage{soul}
\usepackage{amsmath}
\usepackage{amssymb}
\usepackage{pifont}
\newcommand{\cmark}{\ding{51}}%
\newcommand{\xmark}{\ding{55}}%
\usepackage{slashbox}

\renewcommand{\paragraph}[1]{\textbf{#1} }
\newcommand{\x}{$\times$}

\def\BibTeX{{\rm B\kern-.05em{\sc i\kern-.025em b}\kern-.08em
    T\kern-.1667em\lower.7ex\hbox{E}\kern-.125emX}}

\newcommand{\YES}{\textcolor{green}{\cmark}}
\newcommand{\NO}{\textcolor{red}{\xmark}}

\newcommand\blfootnote[1]{%
  \begingroup
  \renewcommand\thefootnote{}\footnote{#1}%
  \addtocounter{footnote}{-1}%
  \endgroup
}

\fancypagestyle{firstpage}{
  \fancyhf{}

  \fancyfoot[C]{\thepage}
}

\title{Sparse Systolic Tensor Array \\ for Efficient CNN Hardware Acceleration} 

\begin{document}

\author{Zhi-Gang Liu*}   
\author{Paul N. Whatmough*}
\author{Matthew Mattina} 

\affil{Arm ML Research Lab, Boston, MA, USA}

\maketitle

\thispagestyle{firstpage}
\pagestyle{plain}


\input{paper}


\bibliographystyle{IEEEtranS}
\bibliography{refs}

\end{document}

%% file: paper.tex
\begin{abstract}

Convolutional neural network (CNN) inference on mobile devices demands efficient hardware acceleration of low-precision (INT8) general matrix multiplication (GEMM).
Exploiting data sparsity is a common approach to further accelerate GEMM for CNN inference, and in particular, structural sparsity has the advantages of predictable load balancing and very low index overhead.
In this paper, we address a key architectural challenge with structural sparsity: how to provide support for a range of sparsity levels while maintaining high utilization of the hardware.
We describe a time unrolled formulation of variable density-bound block (VDBB) sparsity that allows for a configurable number of non-zero elements per block, at constant utilization.
We then describe a systolic array microarchitecture that implements this scheme, with two data reuse optimizations.
Firstly, we increase reuse in both operands and partial products by increasing the number of MACs per PE.
Secondly, we introduce a novel approach of moving the IM2COL transform into the hardware, which allows us to achive a 3\x data bandwidth expansion just before the operands are consumed by the datapath, reducing the SRAM power consumption.


The optimizations for weight sparsity, activation sparsity and data reuse are all interrelated and therefore the optimal combination is not obvious.
Therefore, we perform an design space evaluation to find the pareto optimal design characteristics.
The resulting design achieves 16.8 TOPS/W in 16nm with modest 50\% model sparsity and scales with model sparsity up to 55.7 TOPS/W at 87.5\%.
As well as successfully demonstrating the variable DBB technique, this result significantly out performs previously reported sparse CNN accelerators.

\end{abstract}

\section{Introduction}

\blfootnote{* authors with equal contribution.}

Convolutional neural network (CNN) inference has quickly become an important workload on IoT~\cite{kodali-iccd17,fedorov2019sparse,tinylstm} and mobile computing devices~\cite{euphrates,asv}, which has spurred the development of hardware accelerators in mobile SoCs~\cite{smiv,hansen-icpr20}.
CNNs are fundamentally composed of many layers of multi-channel 2D convolution, interspersed with non-linear activation functions.
The convolutions are usually \textit{lowered} in the runtime to general matrix multiplication (GEMM) by linearizing the 3D feature maps into a 2D structure using the \textit{IM2COL} function~\cite{warden_gemm}.
The resulting GEMMs are usually compute-bound, $O(N^{3})$, and heavily dominate the runtime of CNN inference.
Therefore, GEMM is an obvious target for acceleration~\cite{interstellar-asplos20},
and being compute bound, the speedup justifies the extra silicon real estate.
For mobile computing devices, INT8 CNN inference accelerators demand high energy efficiency (TOPS/W) and area efficiency (TOPS/mm$^2$) to achieve performance and price differentiation.

\begin{figure}[t]
\centering
\includegraphics[width=0.45\textwidth]{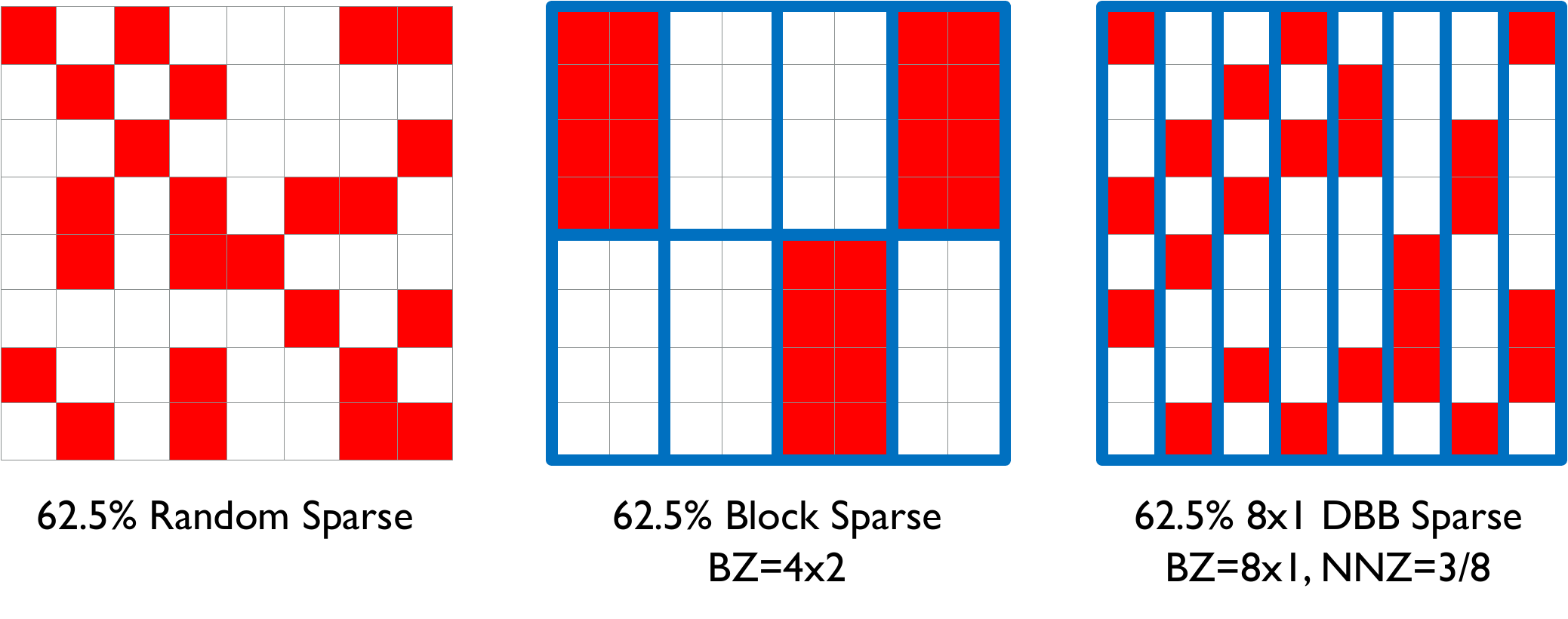}
\\
\vspace{-5pt}
(a)\hspace{70pt}(b)\hspace{70pt}(c)
\caption{Sparse matrix encodings, red denotes non-zero element.
BZ is block size, and NNZ is non-zero elements per block.
}
\vspace{-10pt}
\label{fig:sparse_mats}
\end{figure}

Data sparsity can be exploited in CNN inference accelerators~\cite{eie,scnn,laconic_isca19}, as zeros in the data reduce the theoretical compute and storage cost significantly.
However, traditional sparse matrix multiplication (sGEMM) from scientific workloads only generates speedup at very high sparsity (e.g., $>$95\% zeros).
In contrast, CNNs typically exhibit 50--70\% zeros~\cite{deepcompression,scnn}, which falls well below this level.
Furthermore, the zeros in CNNs are naturally distributed randomly (Fig.~\ref{fig:sparse_mats}(a)), which leads to two challenges.
Firstly, this requires each non-zero element to be indexed explicitly, which increases the overheads of storing and manipulating the indexes, not to mention the cost of gathering and scattering the data.
Indexing overheads are exacerbated for INT8 data types, where the index itself may require at least 4-bits.
Secondly, load balancing is intractable for random sparsity without inspecting the indexes at runtime, making it difficult to achieve high utilization.


Block sparsity (Fig.~\ref{fig:sparse_mats}(b)) is an alternative, where a contiguous block or pattern of elements is forced to be either all-zero or unconstrained.
For hardware design, this has two huge advantages over random sparsity.
Firstly, the indexing cost is substantially reduced, because a single index is amortized over multiple data elements.
Secondly, as the granularity is now a block rather than a single element, the workload is entirely predictable, which makes load balancing and hardware design in general much easier.
However, as the block size increases to the benefit of the hardware, the CNN accuracy degrades due to the increasingly large ``holes'' in the weight matrices.
A fairly new alternative formulation called \textit{density bound block (DBB)} sparsity~\cite{kang-tcsvt19,liu-cal20}, introduces a bound on the number of non-zero elements in each block (Fig.~\ref{fig:sparse_mats}(c)).
DBB sparsity exhibits the hardware advantages of block sparsity and the CNN performance of random sparsity.
DBB has even been implemented in the recently announced Nvidia A100 GPU product~\cite{nv-a100-datasheet}, which cites 2$\times$ speedup for 50\% sparsity.

A significant limitation of the two DBB architectures published to date, is that the sparsity is fixed at design time: 75\% in~\cite{kang-tcsvt19}, and 50\% in~\cite{nv-a100-datasheet}.
As a result, any model that does not meet this fixed sparsity level will be forced to fall back to dense execution with no gains.
Even worse, any models that achieve even \textit{higher} sparsity than the fixed level will also see no further gains.
Therefore, a fixed sparsity level severely limits the usefulness of DBB for broader deployment, as any commercial products are forced to choose a modest sparsity level to best suit the average customer.
The challenge in supporting variable sparsity levels is that the number of MACs required per fixed amount of weights read from SRAM changes.
For a fixed provisioning of hardware MACs, this leads to a proportional drop in utilization, which directly impacts energy efficiency and area efficiency.

In this paper, we introduce a novel variable DBB technique using a time unrolled architecture.
We demonstrate this in a reuse optimized accelerator and demonstrate state-of-the-art results.
The contributions of this paper are summarized below:

\begin{itemize}
\item{
\paragraph{Variable Density Bound Block (VDBB)}
In previous work, the DBB compression is fixed~\cite{kang-tcsvt19,nv-a100-datasheet}, which is a big impediment to broader deployment, because CNNs can vary widely in their weight sparsity.
This paper describes a variable DBB (VDBB) architecture achieved through \textit{time unrolling}, that supports all structured sparsity ratios from 12.5\% ($1/8$) up to fully dense ($8/8$), achieving both speedup and energy efficiency as sparsity increases.
}

\item{
\paragraph{Reuse Optimized VDBB Microarchitecture}
We describe an accelerator microarchitecture to implement time unrolled variable DBB.
At the datapath array level, we implement a systolic tensor array (STA) composed of a more complex PE called a tensor PE (TPE), which increases reuse and better amortizes the cost of data movement.
To decrease SRAM read power we introduce a novel hardware IM2COL unit after the SRAM and just before the datapath, which achieves ~3x bandwidth magnification for 3\x 3 kernels.
We show how to incorporate VDBB weight sparsity and random activation sparsity.
}

\item{
\paragraph{Design Space and Evaluation}
The combination of time unrolled VDBB and the reuse optimized implementation result in a large number of parameters, which all have an interlinked impact on performance, area and energy such that the optimal design point is not obvious.
Therefore, we finally enumerate the design space and describe the pareto-optimal design choices.
Results in 16nm for INT8 at 1GHz show the optimal nominal 4 TOPS accelerator has effective throughput and energy that scale strongly with model sparsity and demonstrating 16.8 TOPS/W (50\% model sparsity) up to 55.7 TOPS/W (87.5\% model sparsity).
This is more than 8\x greater energy efficiency compared to the previously published Laconic~\cite{laconic_isca19}.
}

\end{itemize}

The remainder of the paper is organized as follows.
Section~\ref{sec:background} provides background material on DBB and presents motivation for the paper.
Section~\ref{sec:vdbb} describes the time unrolled variable DBB (VDBB) architecture, and Section~\ref{sec:npu} presents a reuse optimized accelerator implementation.
Section~\ref{sec:method} describes the experimental methodology and Section~\ref{sec:results} presents the results.
Section~\ref{sec:related} describes related work.
Finally, Section~\ref{sec:conclusion} concludes the paper.

\section{Background and Motivation}
\label{sec:background}


The main advantage with DBB weight compression for hardware deployment is that it maintains the regularity of GEMM.
This results in speedup proportional to the compression rate, high utilization, and low index storage and manipulation overheads which is anyway amortized over the block size.
In this section we give an overview of the DBB approach and discuss preliminaries.
We also explain the limitations of the fixed sparsity ratio used in the previous work.

\begin{figure}[!t]
\centering
\includegraphics[width=0.52\textwidth]{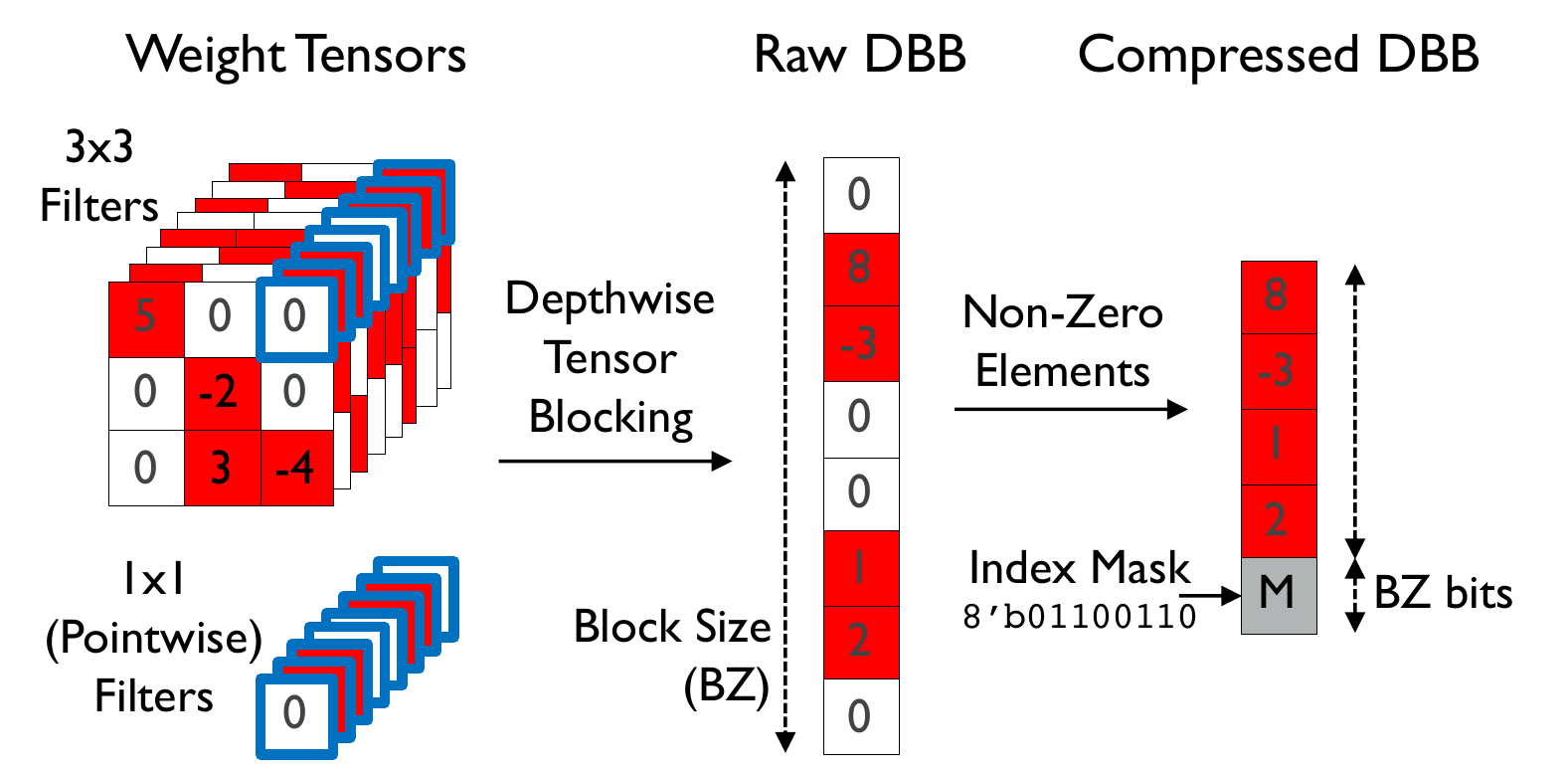}
\caption{
Density Bound Block (DBB) structured sparsity constraints result in a maximum of NNZ non-zero values per block of size BZ, when the weight tensors (e.g., 3\x3, 1\x1 etc) are decomposed in the depth (channel) dimension.
The block is compressed simply by removing the zero elements and appending the index bitmask M, which indicates the location of non-zero elements in the expanded block.
}
\label{fig:dbb_compress}
\end{figure}

\subsection{DBB Overview}

Both the weights and activations of CNNs exhibit sparsity.
However, while the weights are known in advance and can be influenced during training, activations depend on the input image and therefore their sparsity is more difficult to influence.
Therefore, in this work, we apply DBB to weight tensors.
On top of this, we describe clock-gating schemes to exploit activation sparsity.
Density bound block~\cite{kang-tcsvt19} imposes a simple constraint on the sparsity of a block of BZ elements, such that there are at most NNZ non-zero elements per block.
Fig. ~\ref{fig:dbb_compress} gives a concrete example, using a block size of 8\x1.
The tensor blocking is performed depthwise (i.e. over the channel dimension), such that the elements in any single 3\x3 kernel do not fall into the same block, which avoids over-constraining any single kernel.
Note that this approach also works for the 1\x1 (pointwise) filters that represent the majority of the compute in depthwise separable layers~\cite{xception}, which are used heavily in the influential MobileNets~\cite{mobilenetv1_model} family of models.

When the tensors are blocked in this fashion, they can be trivially compressed in two steps.
First, the non-zero elements are stored by removing the zeros.
Secondly, a simple bitmask index M is added to encode the presence of a non-zero element at each location in the expanded block (size BZ).
The resulting compressed size is $8NNZ + BZ$ bits, assuming INT8 word size, giving a compression ratio of $8BZ / (8NNZ+BZ)$.
Any blocks that have less than NNZ non-zero elements will include one or more zero elements in the encoded form.



\subsection{Training DBB CNN Models}
\label{sec:dbb-result}

\begin{table}[!t]
\centering
\scriptsize
\begin{tabular}{l c c c c c}
\toprule
Model                                & Dataset           & Baseline    &  \multicolumn{3}{c}{---------- With DBB Pruning ----------}  \\
                                     &                   & Acc.(\%)    & Acc.(\%)  &  Total NNZ & Sparsity$^1$ (\%)         \\
\midrule                            
LeNet-5                        & MNIST             & 99.1        &  98.7      & 1.05K & 75 ($2/8$)   \\ 
ConvNet                        & CIFAR10           & 86.0        &  85.3      & 26.8K & 75 ($2/8$)   \\
ResNet-50V1                    & ImageNet          & 75.2        &  74.2      & 8.79M & 62.5 ($3/8$) \\
VGG-16                         & ImageNet          & 71.5        &  71.4      & 5.39M & 62.5 ($3/8$) \\
MobileNetV1                    & ImageNet          & 70.9        &  69.8      & 1.6M  & 50 ($4/8$)   \\
\bottomrule
\end{tabular}
$^1$Convolution layers only.
\caption{
CNNs trained with INT8 DBB weights with a block size of 8. 
The maximum block sparsity achieved for these benchmark models varies from 50\% ($4/8$) to 75\% ($2/8$).}
\label{tab:dbb-training}
\end{table}


CNNs must be specially trained to meet the DBB constraint.  
To demonstrate the feasibility of this, we trained five CNNs, applying conventional INT8 quantization and magnitude-based DBB pruning to VGG-16, MobileNetV1, ResNet-50V1, 5-layer ConvNet and LeNet-5 on ImageNet, CIFAR10 and MNIST datasets. 
The DBB sparsity hyperparameter was optimized for each model.
For MobileNet, we apply DBB to the pointwise (1\x1 kernel) layers only, which anyway constitute the vast majority of the ops. For the depthwise separable layers, we fall back to dense operation, which is a key feature of this work. 
The training results are given in Table~\ref{tab:dbb-training}.
Further details of the training methodology are given in Section~\ref{sec:method:training}.
The accuracy loss from combined DBB pruning and INT8 quantization is 0.1--1.1\% across all five models, which include both a relatively big network (ResNet-50V1) and a compact parameter-efficient network (MobileNetV1), both of which are typically tough test cases for model optimizations.
These training experiments validate that DBB pruning achieves 2--4$\times$ weight compression, while maintaining reasonable test accuracy with INT8 quantization, while maintaining regularity.

\subsection{DBB Parameters}

There are only two key parameters for DBB: the block size (BZ) and the density bound given by $NNZ/BZ$\footnote{In this paper we routinely refer to the block \textit{density} as a ratio of NNZ/BZ, but also use the \textit{sparsity} given as a percentage.}.
In general, a larger block size introduces a less severe constraint on the optimization process, but increases the hardware cost.
A larger block size also provides a greater granularity of sparsity levels.
To illustrate this, Table~\ref{tab:dbb_size} shows the training sensitivity to the block size (BZ) using 8-bit quantized LeNet-5 on the MNIST dataset, following the methodology in Section~\ref{sec:method:training}.
For a given sparsity ratio, DBB pruned LeNet-5 models with larger block sizes clearly achieve better predication accuracy.
For example, the ratio of $1/4$ (NNZ=1 and BZ=4) achieves 98.7\% accuracy, but the same compression ratio with a higher block size, e.g. $4/16$, gives better accuracy (99.1\%). 
Based on this, we use a block size of 8 based on the results of the DBB pruning in Table ~\ref{tab:dbb-training} and analysis of the hardware cost.
Previous work uses a block size of 8 in~\cite{kang-tcsvt19} and 4 in~\cite{nv-a100-datasheet}.
Note that any models trained with block size of 4 are guaranteed to be supported on hardware with a block size of 8, as a block 4 model will always satisfy the same sparsity ratio in block 8 format.

\begin{table}[t]
\centering
\begin{tabular}{c | c c c c }
\toprule
\backslashbox{NNZ}{BZ}
 & 2 & 4 & 8 & 16 \\
\midrule 
1 &\cellcolor{green!10}99.0\% & \cellcolor{blue!10}98.7\% & \cellcolor{red!10}98.2\% & \cellcolor{yellow!10}97.9\%   \\
2 &   --    & \cellcolor{green!10}99.1\% &	\cellcolor{blue!10}98.9\% &	\cellcolor{red!10}98.6\% \\
4 &   --    & --      & \cellcolor{green!10}99.1\% & \cellcolor{blue!10}99.1\%
\\ 
\bottomrule
\end{tabular}
\caption{
Accuracy sensitivity to DBB block size (BZ) and number of non-zeros (NNZ) for 8-bit LeNet-5 on MNIST.
Accuracy increases with block size at equal sparsity ratio.
Cell colors indicate equal compression ratios of NNZ/BZ.
}
\vspace{10pt}
\label{tab:dbb_size}
\end{table}

\begin{figure*}[t]
\centering
\hspace{-18pt}
\includegraphics[width=2.1\columnwidth]{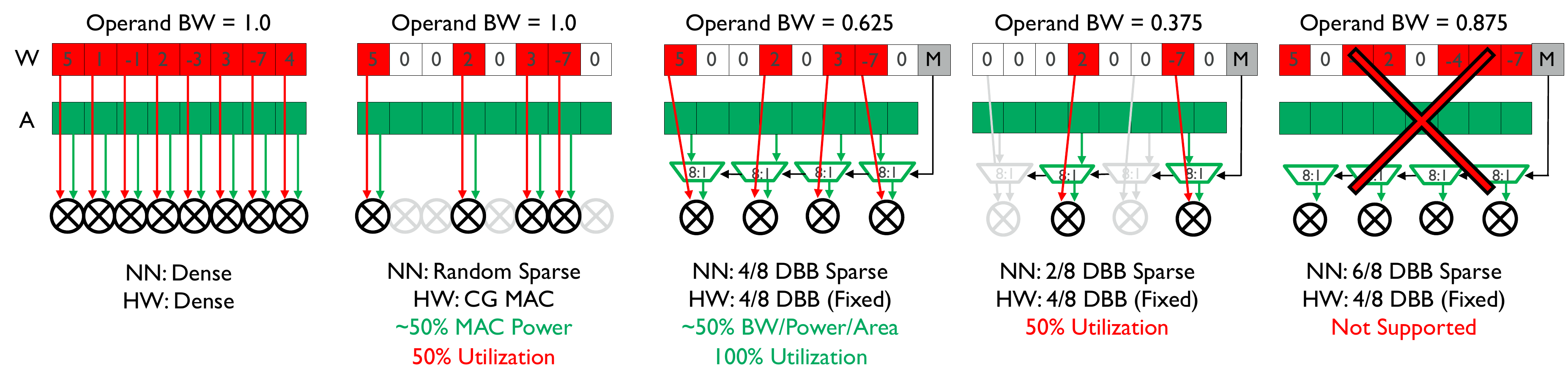}
\\
\vspace{-6pt}
\hspace{-10pt}(a)
\hspace{89pt}(b)
\hspace{89pt}(c)
\hspace{89pt}(d)
\hspace{89pt}(e)
\caption{
Spatially unrolled datapaths, all with effective throughput of 16 Ops/cycle. 
(a) Conventional dense datapath with no benefit from sparsity. 
(b) With a random sparsity, we can clock gate (CG) on zero operands to proportionally reduce compute power while lowering utilization.
However, this does not reduce data movement power or area.
(c) A DBB datapath designed for $4/8$ block sparsity has the same effective throughput as (a), but requires 62.5\% operand bandwidth, and about half the area/power.
The block sparsity ($4/8$) is fixed at silicon design time.
(d) A model with higher sparsity ($2/8$) has little advantage, as the hardware is designed for ($4/8$) block sparsity.
(e) A model with lower sparsity (e.g., $6/8$) is not natively supported.
}
\label{fig:fixed-dbb}
\end{figure*}

\subsection{Motivation}

We showed in Table~\ref{tab:dbb-training} that popular CNN architectures achieve a fairly wide range of DBB pruning ratios, which vary depending on the dataset, the network architecture and even the training recipe.  
Smaller models such as LeNet-5 and ConvNet can be pruned down to $1/4$ of their original size, so would ideally benefit from a block sparsity rate of $2/8$.  
However, very compact models such as MobileNet are notoriously tricky to train and optimize and achieve about 50\% compression at best, which requires a block size of $4/8$.
We are also very likely to encounter a variety of sparsity levels within a single model.
For example, it is very common to avoid optimizing the first layer of a large CNN, as this often damages accuracy.
It is also possible to optimize sparsity per-layer or even per-channel to extract the most from the model.
\textit{Therefore, all of this points towards the need to support a range of structured sparsity ratios natively in the hardware.}

Previous implementations of DBB have demonstrated fixed sparsity: Kang et al. implements fixed DBB with a $2/8$ (75\%) block, and the Nvidia A100 GPU~\cite{nv-a100-datasheet} implements a $2/4$ block.
However, the fixed block sparsity ratios are a practical limitation, as models with more sparsity will see no further improvement, and models with less sparsity will have to fall back to dense GEMM.
In this work, we demonstrate an effective approach to implement DBB with continuously variable block sparsity.
We also demonstrate the first structured sparsity systolic array, which heavily emphasizes data reuse.

\section{Variable Density Bound Block (VDBB)}
\label{sec:vdbb}

As we outlined in the previous section, hardware support for variable sparsity DBB (VDBB) is highly desirable. 
However, varying the density bound leads to hardware under utilization.
In this section we will discuss this issue more concretely and present an architecture solution to the VDBB requirement.


\subsection{Spatially Unrolled Block Architecture}

A conventional (dense) datapath is shown in Fig.~\ref{fig:fixed-dbb}(a), where a block of 8 weights (W) are multiplied by a corresponding block of activations (A).
The most obvious approach to compute a sparse block is to parallelize the operations across independent hardware MACs, i.e. spatially unroll the block.
For random weight sparsity, we can add a simple mechanism to detect zero operands and clock gate (CG) the relevant MAC lane~\cite{eyerissIsca,minerva}, as shown in Fig.~\ref{fig:fixed-dbb}(b).
This reduces the compute power (proportionally to the sparsity), but also reduces the utilization of the hardware, which impacts area efficiency (TOPS/mm$^2$).
It is also challenging to reduce data storage cost with random sparsity, due to the unpredictability of the non-zero elements per fixed SRAM access size.

In contrast, DBB results in a predictable number of non-zero (NNZ) elements per block, which means we can easily reduce both compute and data movement.
For example, Fig.~\ref{fig:fixed-dbb}(c) illustrates a datapath that supports a $4/8$ density-bound block and achieves the same throughput as Fig.~\ref{fig:fixed-dbb}(a).
The DBB version requires only four hardware MACs, each augmented with an 8:1 mux to steer the correct activation element into the MAC.
The select signal for the mux is driven by the positional index metadata ($M$), which is an additional byte per block overhead in this example.
Note that the implementation of DBB by Kang~\cite{kang-tcsvt19} is similar to this, but with a $2/8$ block (75\% sparsity).

However, as we noted in Section~\ref{sec:background}, real world models can exhibit a very wide variety of sparsity levels.
However, the fixed DBB hardware in Fig.~\ref{fig:fixed-dbb}(c) can only support a single fixed block sparsity ratio.
If the sparsity is higher than $4/8$, e.g., $2/8$ shown in Fig.~\ref{fig:fixed-dbb}(d), then the utilization drops, limiting the TOPS/mm$^2$ advantage on sparser models.
Conversely, lower sparsity models, such as the $6/8$ example in Fig.~\ref{fig:fixed-dbb}(e), are not supported and it is necessary to fall back to dense execution, which offers no benefit at all.
Therefore, instead of fixing the sparsity at design time, we would instead like to support all sparsity ratios from very sparse ($1/8$ density) up to fully dense ($8/8$), from 87.5\% to 0\% sparsity.

\subsection{Time Unrolled Block Architecture}
\label{sec:vdbb:time}

The big challenge with supporting variable density bound blocks (VDBB) in hardware, is that as the weight sparsity rate is increased, the hardware utilization decreases, which leads to low energy and area efficiency.
If we implement fixed $4/8$ DBB (50\% DBB compression), a model that achieves $6/8$ would result in a utilization drop of roughly 50\%. 
On top of this, executing a model with sparsity lower than 50\% is not supported, other than by treating it as a dense model.

\begin{figure}[!t]
\centering
\hspace{-5pt}
\includegraphics[width=1\columnwidth]{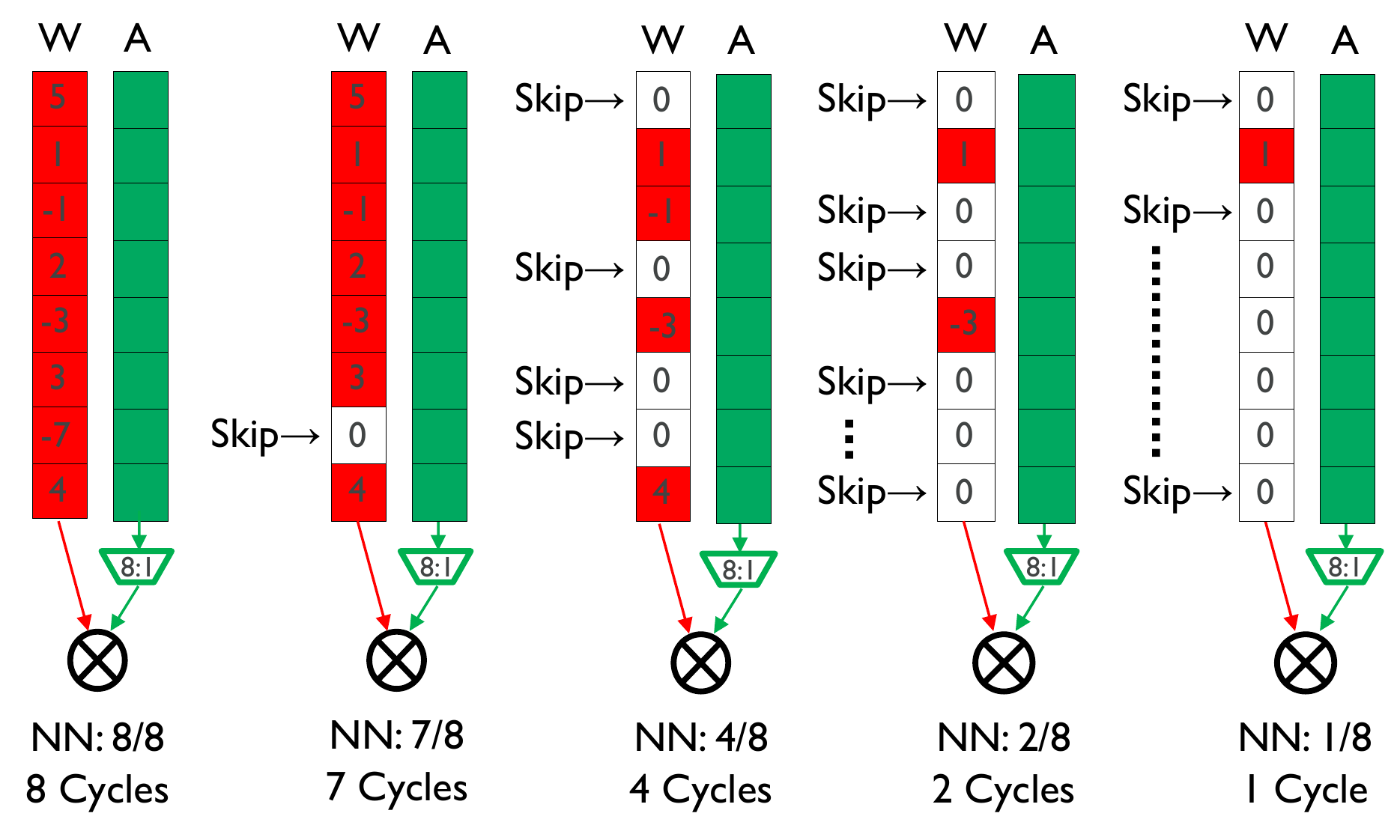}
\\
\vspace{-5pt}
\hspace{0pt}(a)
\hspace{40pt}(b)
\hspace{40pt}(c)
\hspace{40pt}(d)
\hspace{40pt}(e)
\caption{
Time unrolled structured sparse block processing, which allows a continuously variable NNZ per block, while retaining 100\% hardware utilization and constant operand bandwidth.
All NNZ options are supported, from the fully dense case (a), through to 87.5\% sparsity (e).
}
\label{fig:variable-dbb}
\end{figure}

To get around this issue with spatially unrolled DBB architectures~\cite{kang-tcsvt19}, we instead implement the DBB hardware by unrolling the block in the time dimension.
This simply means that we process one element of the density bound block per cycle using a single MAC per block.
Of course, the key advantage of this arrangement, is that we can now freely vary the block sparsity, while the datapath utilization and the operand bandwidth both remain constant.
The number of cycles per block is the only thing that varies as we change the sparsity, i.e. the effective throughput increases with sparsity.
For example, Fig.~\ref{fig:variable-dbb}(a) shows a conventional dense block unrolled in the time dimension and requiring eight cycles to compute on a single MAC.
While in Fig.~\ref{fig:variable-dbb}(b)--(e), we illustrate that NNZ can be freely varied, with the number of clock cycles required to compute the block being equal to NNZ.
At the extreme, a very sparse model with $1/8$ DBB sparsity only requires 1 cycle per block (8$\times$ speedup).

Although the illustrative diagrams in Fig.~\ref{fig:fixed-dbb} and Fig.~\ref{fig:variable-dbb} show both the zero and non-zero elements of the 8-element block, the key idea with DBB is that the data can be trivially compressed (Section~\ref{sec:background}), by storing only the non-zero elements and the index metadata M.
Therefore, the non-zero elements of the weight block are consumed one per cycle, and the skipping of elements is achieved implicitly.
The corresponding activation element is then muxed into the MAC.
Note that a complex reordering buffer is not required to implement this, and it results in very high energy and area efficiency.


\section{Accelerator}
\label{sec:npu}

This section describes an extremely efficient VDBB implementation, that aggressively optimizes five types of data reuse.
The proposed accelerator (Fig.~\ref{fig:system}) includes a Systolic Tensor Array (STA), local SRAMs for weights and activations, IM2COL activation bandwidth magnifier, and Arm Cortex-M33 microcontrollers for DMA and vector compute operations.

\subsection{Datapath Array}
\label{sec:npu:sa}




The systolic array (SA) is a very efficient and scalable hardware implementation of GEMM, due to the local register-to-register operand reuse.
However, implementing VDBB in an systolic architecture greatly improves energy and area efficiency.
We achieve this by generalizing the SA into the systolic tensor array (STA), which is amenable to supporting DBB and VDBB.




\subsubsection{\textbf{Systolic Tensor Array (STA)}}
The classic systolic array (SA) of Fig.~\subref*{fig:array:sa}, consists of an M\x N array of PEs.
Each PE is a single MAC with INT8 operand (OPR) pipeline registers a INT32 accumulator (ACC) register.
We use an output stationary dataflow, which allows the larger INT32 accumulators to remain stationary.
The Systolic Tensor Array (STA) of Fig.~\ref{fig:array:sta} extends the SA concept, with a more complex PE called a tensor PE (TPE).
The TPE accepts a tensor (matrix) of weights and a tensor of activations per cycle, instead of a single weight and activation.
Instead of computing a single MAC per cycle, each TPE essentially processes a small matrix multiplication on the input matrices of size A\x C, using a B-way dotproduct (DP).
This increases the MACs to operands ratio, which we refer to as \textit{intra-TPE reuse}.
While moving from a scalar MAC to a dot product introduces \textit{accumulator reuse}.
In the remainder of this paper, we uniquely denote an STA configuration\footnote{The classic SA (Fig.~\ref{fig:array:sa}) is a special case: 1$\times$1$\times$1\_M$\times$N.
Dot product architectures similar to DaDianNao~\cite{dadiannao2014micro} can be described as 1\x B\x1\_1\x1.} 
 as A$\times$B$\times$C\_M$\times$N.
Fig~\ref{fig:dataflow:dbb} illustrates the STA dataflow, which is similar to the classic SA, but with tensor (i.e. sub-matrix) operands in place of scalar operands.

\begin{figure}[!t]
\centering
\includegraphics[width=0.8\columnwidth]{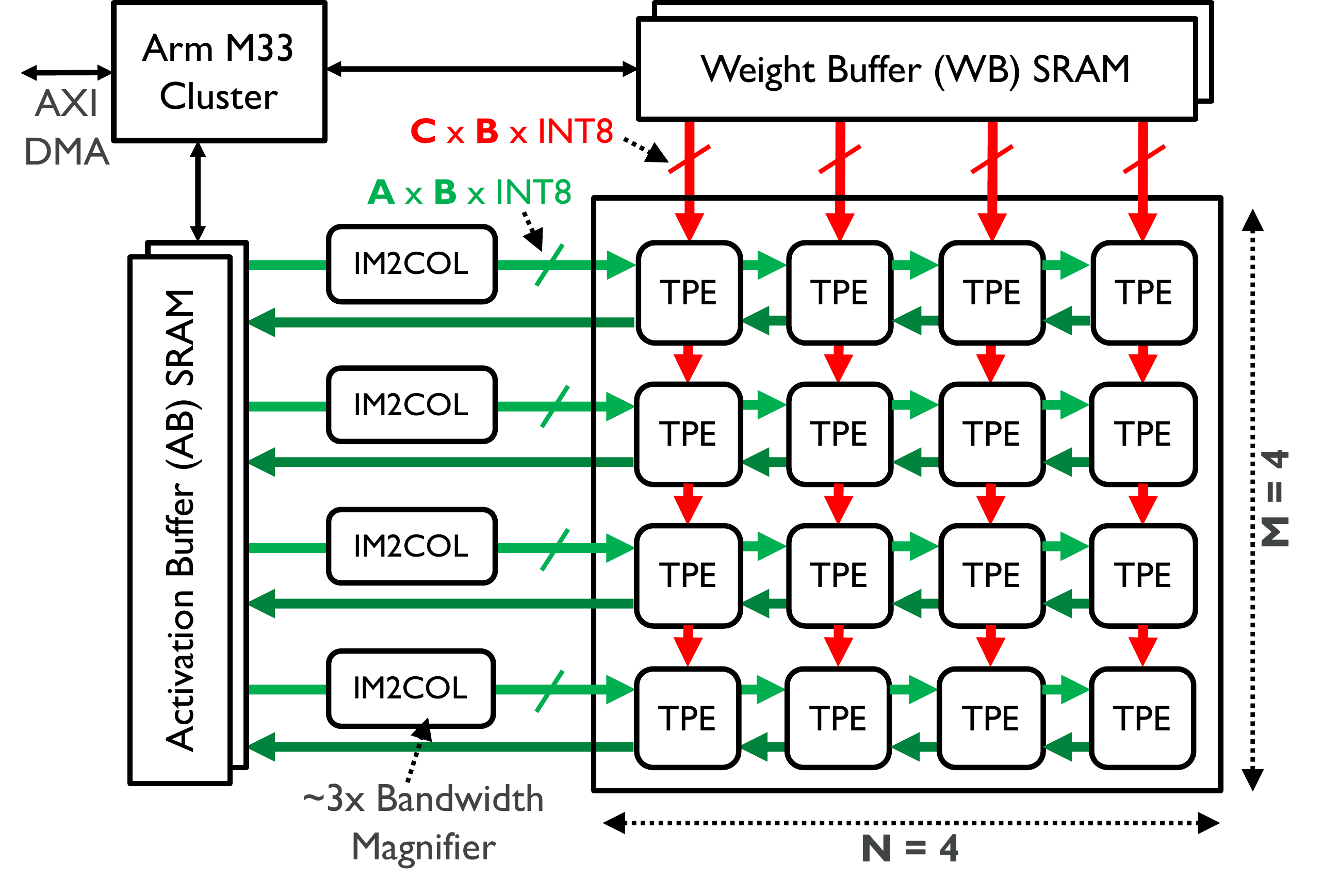}
\caption{VDBB accelerator microarchitecture consisting of TPE datapath array, local SRAM, IM2COL unit, and M33 MCUs. }
\label{fig:system}
\end{figure}



\subsubsection{\textbf{Adding DBB Support (STA-DBB)}}
\label{sec:npu:dbb}
\begin{figure*}[!t]
    \centering
    \subfloat[Systolic Array\newline(SA) 1$\times$1$\times$1\_4$\times$4]{
        \centering
        \includegraphics[width=0.50\columnwidth]{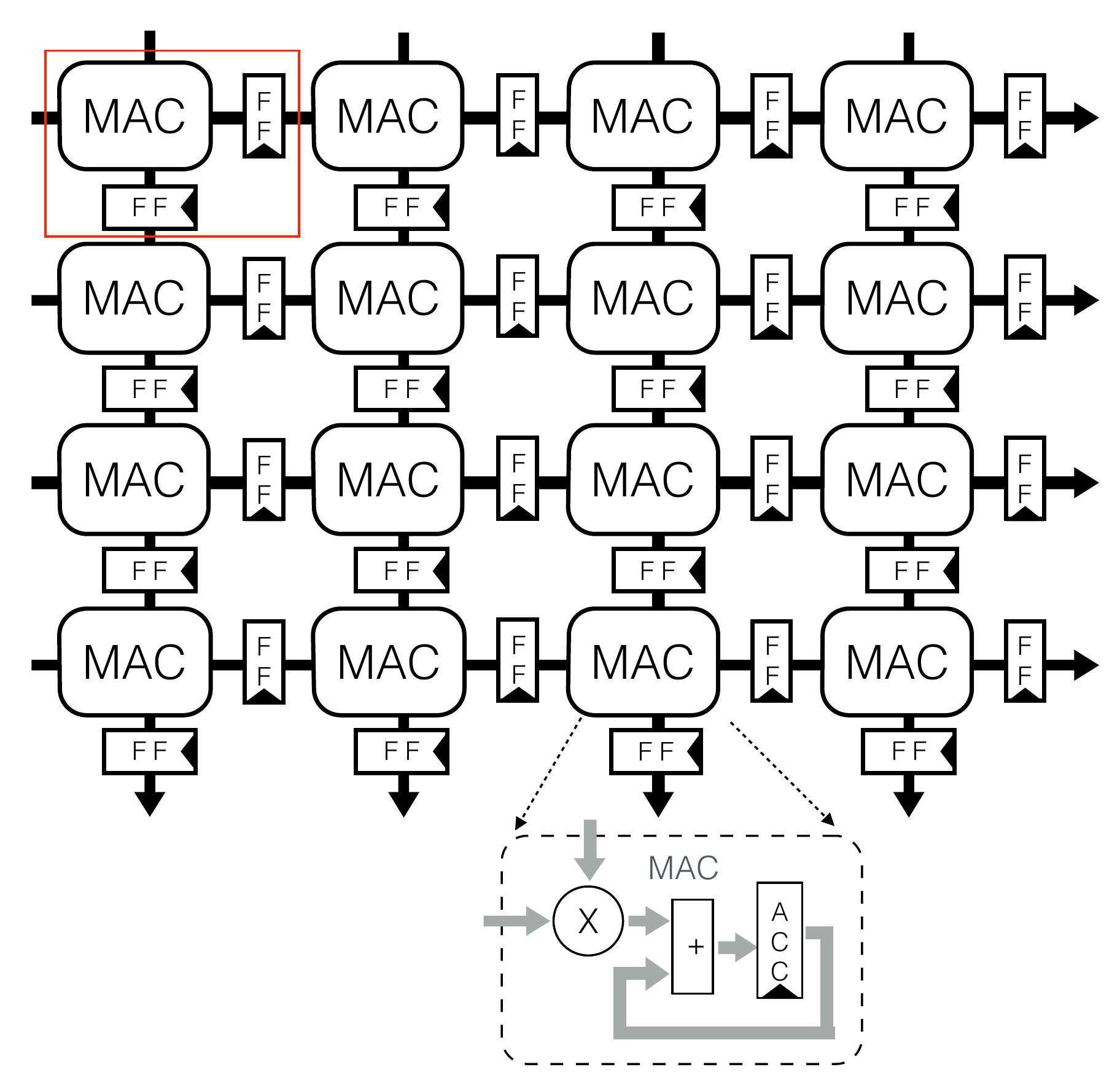}
        \label{fig:array:sa}
    }
    \centering
    \hspace{1pt}
    \subfloat[][Systolic Tensor Array\newline (STA) 2$\times$4$\times$2\_2$\times$2]{
        \centering
        \includegraphics[width=0.45\columnwidth]{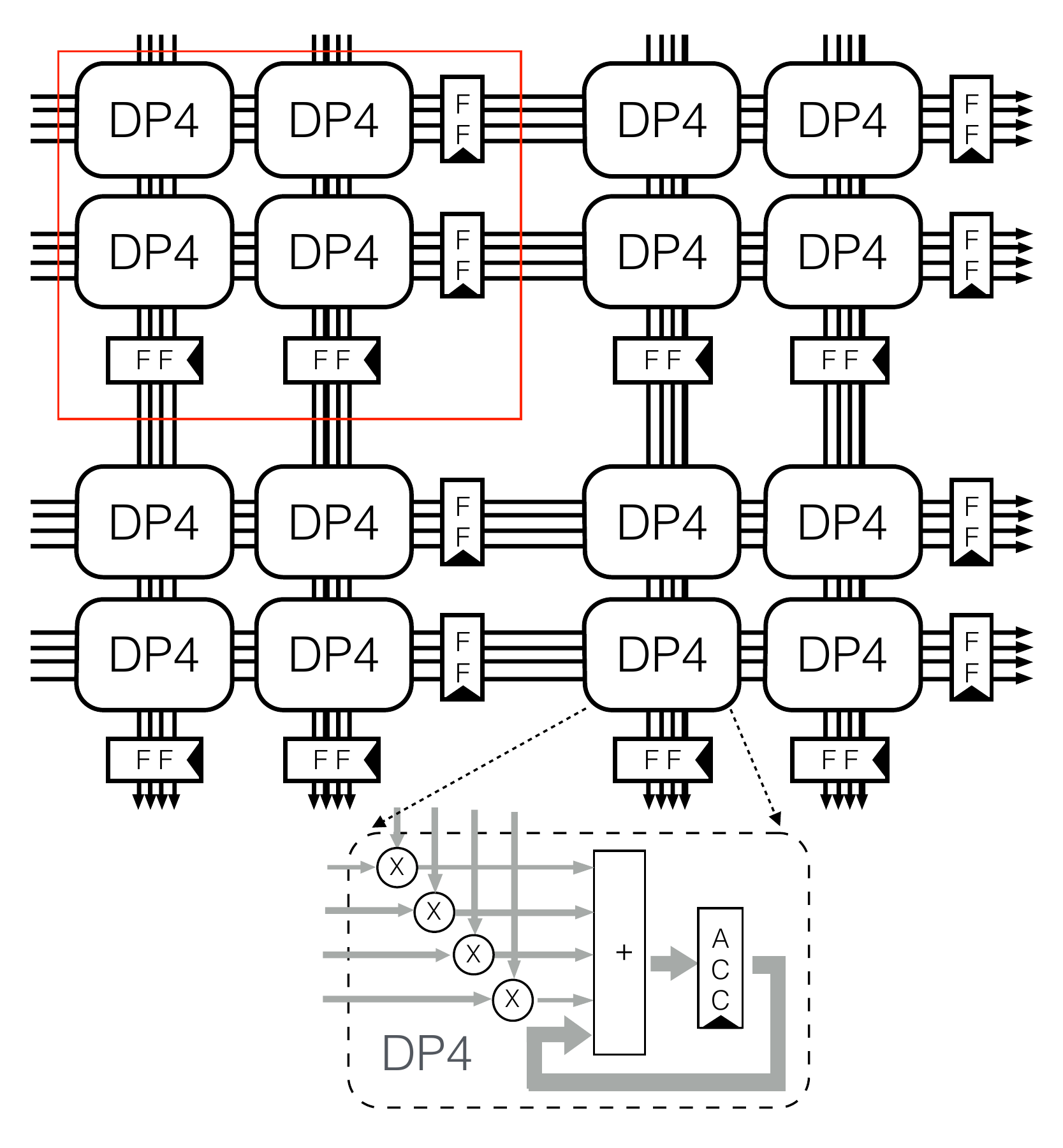}
        \label{fig:array:sta}
    }
    \hspace{1pt}
    \subfloat[][STA with DBB support\newline(STA-DBB) 2$\times$4$\times$2\_2$\times$2]{
        \centering
        \includegraphics[width=0.46\columnwidth]{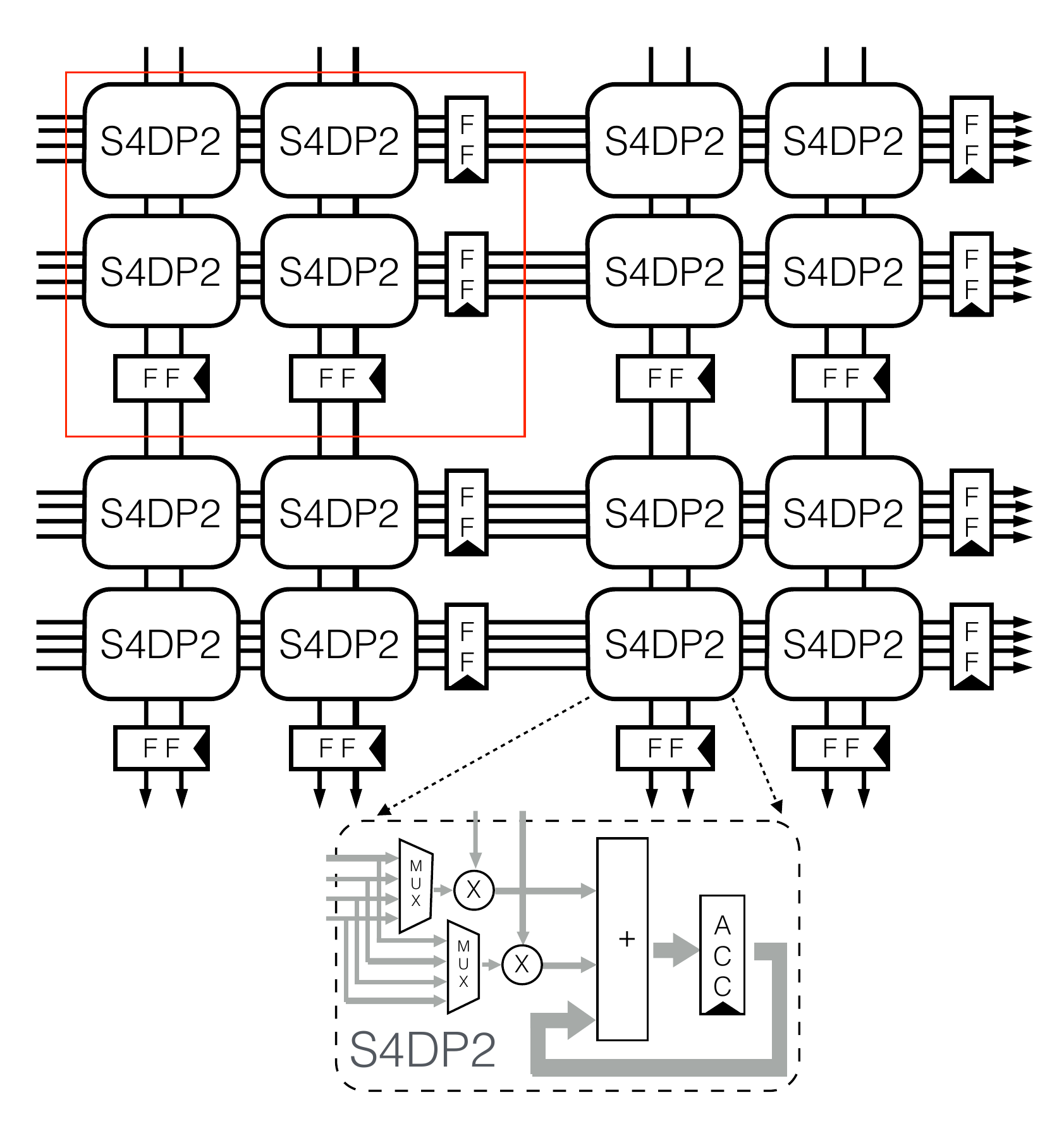}
        \label{fig:array:sta-dbb}
    }
    \hspace{1pt}
    \subfloat[][STA with VDBB support\newline (STA-VDBB) 2$\times$8$\times$4\_2$\times$2]{
        \centering
        \includegraphics[width=0.47\columnwidth]{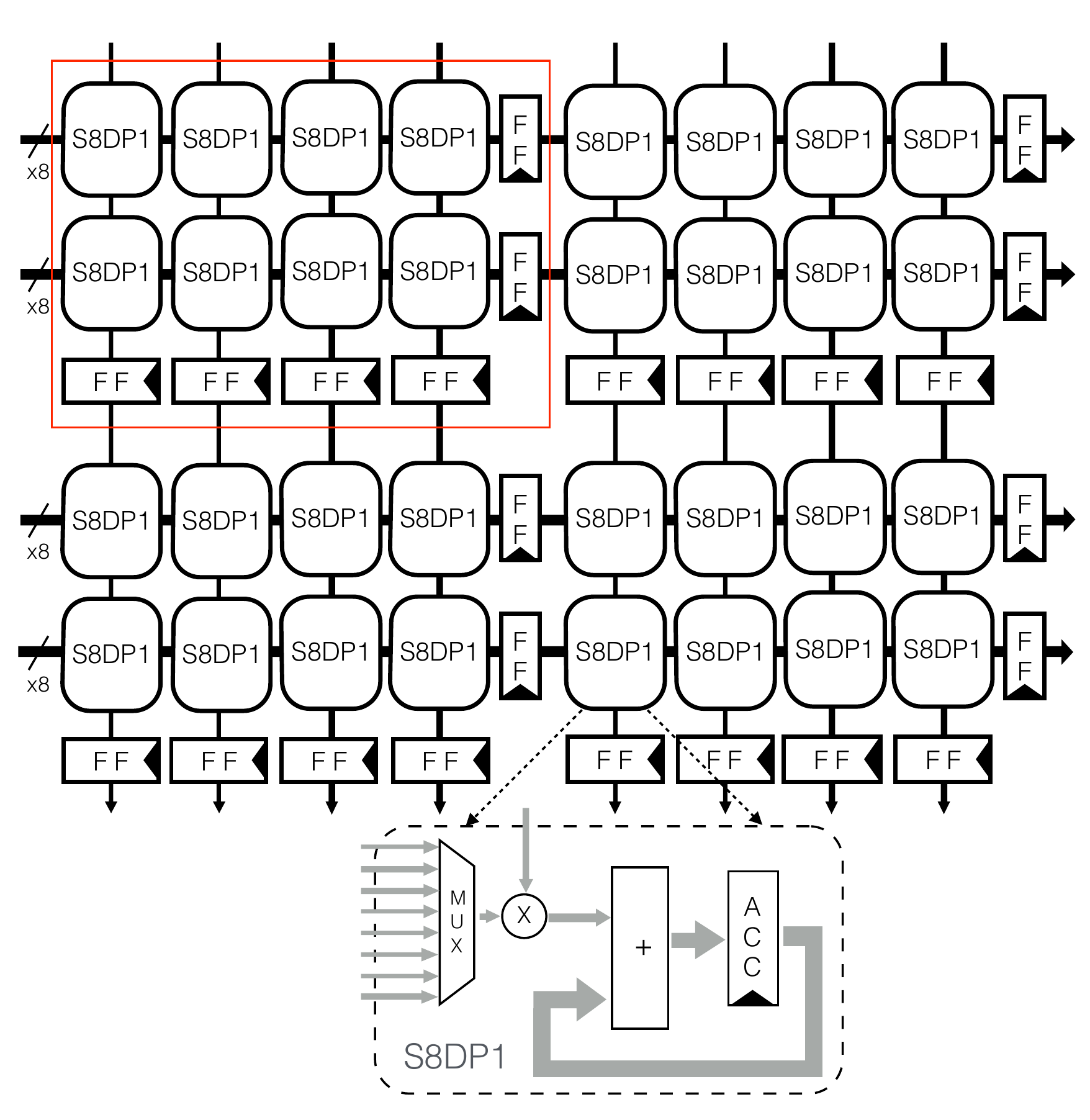}
        \label{fig:array:sta-vdbb}
    }
\caption{
(a) The systolic array (SA) is efficient because operands read from SRAM are reused many times as they propagate through PEs in the M\x N array.
(b) The systolic tensor array (STA), generalizes the scalar PE into a tensor PE (TPE), which accepts two tensor operands and performs a small matrix multiplication on each cycle.
This allows us to introduce intra-TPE reuse and accumulator reuse, increasing the ratio of compute to data movement.
(c) Fixed DBB is implemented inside STA by adding a mux to the activation input on the dot product.
(d) Finally, variable DBB is implemented by switching to multiple single MACs to allow time unrolling.
Notation: A$\times$B$\times$C\_M$\times$N denotes a M$\times$N 2-D array of A$\times$B$\times$C TPEs (red box). 
DP2 denotes a 2-way dot-product into a single accumulator register.
S4DP2 denotes a 2-way sparse dot-product (SDP) with a 4:1 mux in the activation path for DBB sparsity.
}
    \label{fig:array}
\end{figure*}

Next, we add support for DBB weight matrices.
DBB allows us to reduce the number of MACs per block from BZ to NNZ, which reduces the area by the compression factor NNZ/BZ.
Each dot product (DP) also requires an additional multiplexer (mux) in front of each MAC to select the activation element that corresponds to the non-zero weight, indicated by the bitmask index, M.
An example 2\x4\x2\_2\x2 STA-DBB configuration is shown in Fig.~\subref*{fig:array:sta-dbb}.
Each TPE is composed of 4\x 4-input 2-way Sparse Dot Product (S4DP2) units, each of which has a 4-value activation vector input $[A_0, A_1,$ $A_2, A_3]$ from the left, and a 2/4 DBB compressed weight input $[W_0, W_1]$ and associated 2-bit non-zero index from the top.
Compared to a conventional (dense) DP4, each S4DP2 trades two MACs for two 8-bit 4:1 datapath multiplexers (MUX), where each MUX costs significantly less than a MAC.
Although this architecture still supports conventional dense GEMM, it only supports a single sparsity ratio (50\% in this example).
Fig.~\subref*{fig:dataflow:dbb} illustrates the corresponding dataflow for this STA-DBB example.
Here, we multiply a 8\x4 activation matrix, A, by a 4\x8 weight matrix with 2/4 DBB sparsity, W.
Both matrices are first partitioned into 2\x4 or 4\x2 sub-matrices respectively, with the W sub-matrix of compressed down to only non-zero elements.
Then each sub-matrix tensor is skewed by one cycle in time across the edges, similar to a conventional SA, but at tensor granularity.    
Finally, each sub-array tensor is input one column (row) per cycle, corresponding with each TPE on the left (top) edge. All of this behaviour is essentially the same as an SA, but replacing scalars with sub-matrices.

\subsubsection{\textbf{Adding VDBB Support (STA-VDBB}}
\begin{figure*}[!t]
    \centering
    \subfloat[Data flow for STA-DBB (2\x4\x2\_2\x2) ]{
        \centering
        \includegraphics[width=0.4\textwidth]{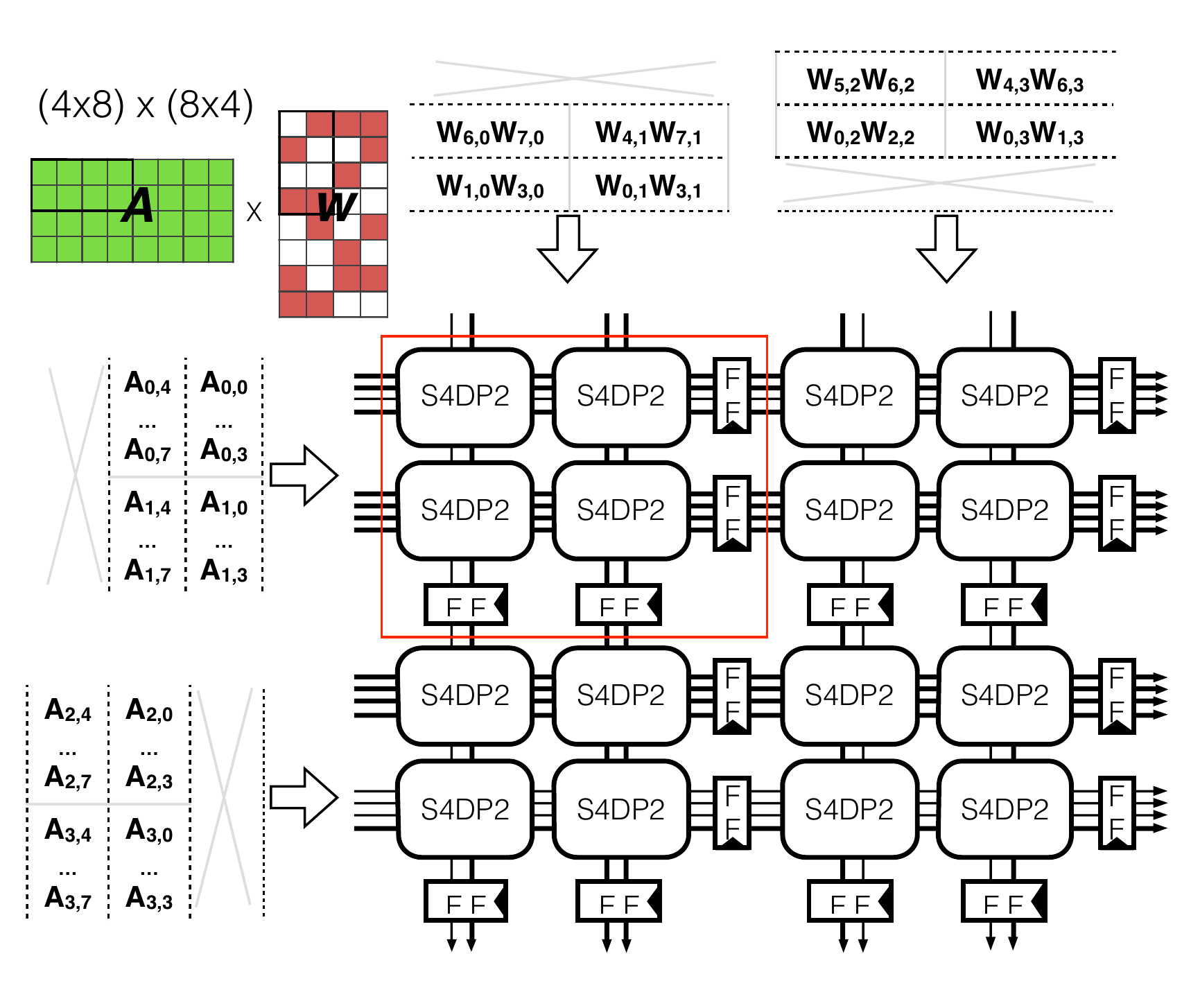}
        \label{fig:dataflow:dbb}
    }
    \hspace{30pt}
    \subfloat[][Data flow for STA-VDBB (2\x8\x4\_2\x2)]{
        \centering
        \includegraphics[width=0.43\textwidth]{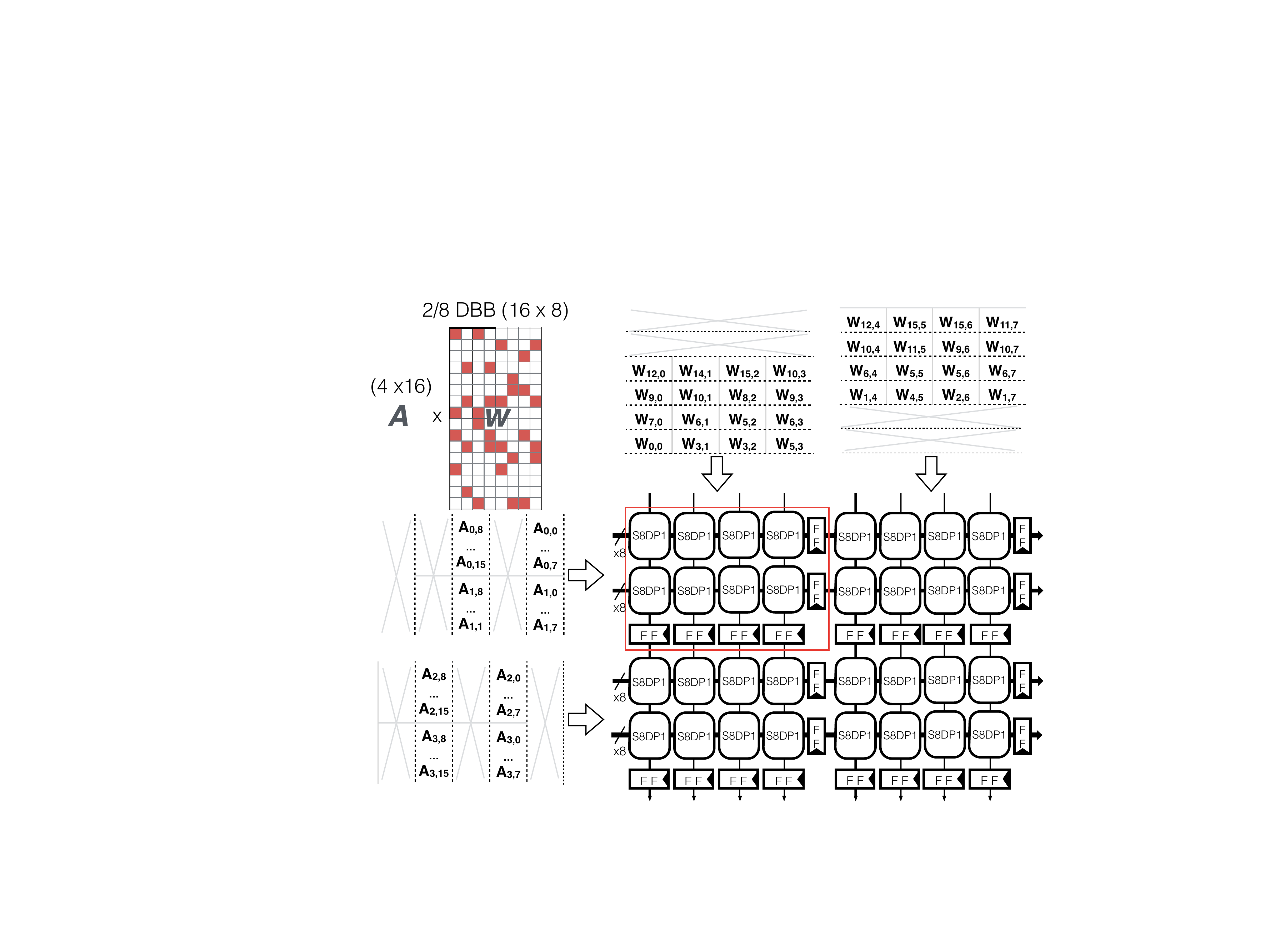}
        \label{fig:dataflow:vdbb}
    }
\caption{
DBB and VDBB dataflow examples on small arrays.  
(a) Example dataflow to compute dense 4\x8 by 2/4 DBB sparse 8\x4 matrix multiply using STA-DBB 2\x2\x4\_2\x2 datapath in 5 clock cycles. 
(b) Example dataflow to compute dense 4\x16 by 2/8 DBB sparse 16\x8 matrix multiply on a STA-VDBB 2\x8\x4\_2\x2 datapath in 8 clock cycles. 
}
\label{fig:dataflow}
\end{figure*}

Finally, we implement time unrolled variable DBB (Section~\ref{sec:vdbb:time}) as an efficient STA-VDBB, by modifying the TPE (Fig.~\subref*{fig:array:sta-vdbb}).
We retain the MUX at the input to the MAC on the activation side to select the required activation according to the bitmask index, $M$. 
But to support VDBB, we move from a wide dot product with accumulator sharing, to a single MAC with an accumulator dedicated to a single block (S8DP1). 
Most importantly, as we are unrolling in time, the occupancy (number of clock cycles) of the S8DP1 unit for each block depends on the compression ratio. 
Fig.~\subref*{fig:dataflow:vdbb} illustrates a corresponding data flow for computing a 4\x16 by 16\x8 matrix multiplication ($A \times W$, respectively),
where $W$ can be compressed into (4\x8) in 2/8 DBB format. 
$A$ and the compressed $W$ are then partitioned into 2\x8 and 2\x2 sub-matrices respectively. 
Each sub-matrix tensor is skewed by one cycle in time across the array edges at the sub-matrix granularity, before being input to one column (row) of the TPE edge.
However, in order to unroll the sparse dot product in each data block, the compressed tenors of $W$ are input one row per clock. 
For each TPE, the corresponding tensor input from the left edge need is delayed until the whole block is complete, i.e. 2 cycle occupancy for this example. 
Due to the DBB sparsity, all TPEs have the the same occupancy in a computing stream. 
For higher DBB compression ratios, the TPE has a lower number of occupancy cycles, resulting in higher throughput, and area/energy efficency.

\begin{table}[!t]
\centering
\scriptsize 
\def\arraystretch{1.5}
\begin{tabular}{l c c c c }
\toprule
Trade-offs                  &    SA        &   STA          &   STA-DBB     &   STA-VDBB    \\
\midrule 
MACs per TPE             & 1         & \tiny$A\times B \times C$ & \tiny $A\times b \times C$ & \tiny $A\times C$   \\
ACCs per TPE             & 1         & \tiny $A\times C$        &  \tiny $A\times C$         & \tiny $A\times C$   \\
OPRs per TPE             & 2         & \tiny $B(A+C)$           & \tiny $A B + b C$            & \tiny $A B + n C$      \\
Inter-TPE Reuse$^1$      & $\frac{M N}{M + N}$ & $\frac{A M C N }{A M + C N}$ & $\frac{A b C M N}{A B M+C b N}$ &  $\frac{A n C M N }{A B M+C n N}$          \\
Intra-TPE Reuse$^2$      & $\frac{1}{2}$  & $\frac{AC}{A+C} $      &  $\frac{A b C}{A B + b C}$   & $\frac{A n C}{A B + n C}$\\     
ACC Reuse         &  1        & $B$     &  $b$             &   1            \\     
A Sparsity CG            & \YES      & \NO     &  \NO            & \YES        \\     
W Sparsity      & \NO       & \NO      & Fixed DBB     & Variable DBB         \\     
\bottomrule

\end{tabular}
\def\arraystretch{1}
\\
\vspace{3pt}
$^1$ Array MACs / Array input operands.
$^2$ TPE MACs / TPE input operands. 
\caption{
Summary of array design trade-offs.  
$b$ indicates the number of MACs in the SDP unit of the STA-DBB datapath. $n$ denotes NNZ for the block.
The STA-VDBB array increases inter- and intra-TPE reuse, while also supporting VDBB weight sparsity and random activation sparsity clock gating (CG).
}
\vspace{10pt}
\label{tab:tpe}
\end{table}

\subsubsection{\textbf{Array Design Trade-offs}}
\label{sec:npu:sa:tradeoff}
Table~\ref{tab:tpe} summarizes the key differences between the conventional SA, the STA, and the sparsity optimized STA-DBB and STA-VDBB designs.
This highlights the analytical benefit we achieve in both inter- and intra-TPE reuse.
However, there are some trade-offs between the items listed, which we touch upon in the summary below:

\begin{itemize}
\item \textbf{Inter-TPE Operand Reuse}
An M\x N systolic array features weight and activation operand reuse, $O(M)$ and $O(N)$ respectively, which amortizes the relatively high cost of reading operands from SRAM at the array edge.

\item \textbf{Intra-TPE Operand Reuse}
STA extends the array-level reuse, by introducing additional operand reuse inside the TPE itself.
This further amortizes data movement, by locally performing a small matrix multiply (with many MACs) on the input tensor operands inside the TPE.

\item \textbf{Accumulator Reuse}
STA introduces accumulator reuse, whenever a wide dot product is used.
Accumulator reuse increases area efficiency by increasing the MACs to registers ratio by using more efficient carry-save implementation in the dot product datapath.

\item \textbf{Weight Structured Sparsity}
To support VDBB in STA, we must switch from wide dot products to single-MACs, sacrificing the area reduction from accumulator reuse.
However, the advantages of VDBB far out this concession.

\item \textbf{Activation Sparsity}
Finally, we can also exploit activation sparsity on top of VDBB weight sparsity, by clock gating (CG) the datapath on zero activations.
This can not be applied to wide dot products, as all inputs would have to be zero, which is very unlikely.
But, we can apply this to STA-VDBB which anyway uses single-MACs.
\end{itemize}

\subsection{Local SRAM}
\label{sec:npu:ram}
As is commonplace for accelerators, we heavily leverage local software managed SRAM~\cite{li-dac19} to provide a low-cost operand supply for the datapath array.
The weight buffer (WB) is 0.5MB and the activation buffer (AB) is 2MB (Fig.~\ref{fig:system}).
Due to the array architecture, the SRAM is grouped together, rather than distributed, so we are able to choose large SRAM instances which fully amortize the cost of the SRAM periphery against the bitcell array.
We further balance the choice of the bank muxing parameter to balance power and area. 
The AB and WB are both double buffered to allow them to be shared between the datapath and the local MCUs.
The input image is loaded into the AB before operation begins, via DMA from the MCUs.

\subsection{IM2COL Unit} 
\label{sec:npu:im2col}

The main drawback of GEMM (compared to native convolution) is the memory footprint overhead from  IM2COL expansion.
IM2COL is used to linearize 3D volumes of CNN feature map and kernel data, in order to process them using GEMM.
If the stride is less than the kernel size, IM2COL results in duplicated pixels in the output.
This leads to an increased storage requirement, and higher SRAM read power.

\begin{figure}[!t]
\centering
\includegraphics[width=0.45\textwidth]{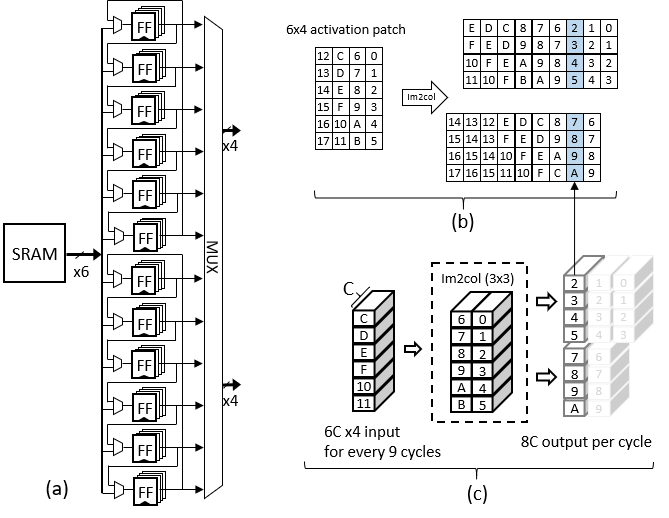}
\caption{
Overview of the hardware IM2COL unit: (a) hardware; (b) IM2COL operation for a 6\x4 activation patch; 
(c) hardware operation.  
The IM2COL unit caches 6$\times$4 inputs in the 6$\times$2 buffers every 9 cycles.
Two 4$\times$ outputs are generated per cycle, reducing SRAM bandwidth by 3$\times$ for a typical 3$\times$3 filter.
}
\label{fig:im2col}
\end{figure}

We directly address this issue by adding a new hardware unit that functions as a SRAM read bandwidth magnifier (Fig.~\ref{fig:system}).
To do this, we implement IM2COL in hardware on activations as late as possible in the microarchitecture: after it is read from local SRAM, and just before the data reaches the datapath (Fig.~\ref{fig:system}).
This allows us to achieve the lower memory footprint of native convolution, while taking advantage of the compute regularity of GEMM, which can be more readily optimized in hardware.
The net result of the late IM2COL hardware unit is a reduction in SRAM read bandwidth for 3$\times$3 and 5$\times$5 convolutions.
This unit consumes data from the SRAM which is held in a small internal buffer register array.
By combining the contents of this buffer, IM2COL transformed outputs are generated.
Fig.~\ref{fig:im2col} illustrates the operation on a 4$\times$6 input feature map tile, which results in 3\x average SRAM read reduction. 


\subsection{Local MCU with SIMD}
\label{sec:npu:mcu}
Although matrix multiplication represents by far the majority of the computation for CNN inference, there are a number of ancillary operators that must be supported to allow the whole model to be processed in place, without moving intermediate results back to the CPU.
These operators include activation functions, pooling, scaling, batch normalization and data type casting.
We implement these in software, using Arm Cortex-M33~\cite{arm_m33} microcontrollers (MCUs).
The M33 MCU has 32-bit SIMD instructions that can encode up to four INT8 operations~\cite{arm_v8m}.
A small 64KB SRAM is included as a program store for the M33, which has minimal impact on the area efficiency.
Control and data movement tasks are also performed by the MCU.
The number of M33 MCUs required depends on the peak throughput, 2 is sufficient for an design with 2 TOPS peak throughput, 4 for 4 TOPS and 8 for 16 TOPS.
The silicon area of the M33 in 16nm technology is very small at 0.008mm$^2$, and the typical power consumption is 3.9$\mu$W/MHz~\cite{arm_m33}.

\begin{figure*}[!t]
    \centering
        \includegraphics[width=2.0
        \columnwidth]{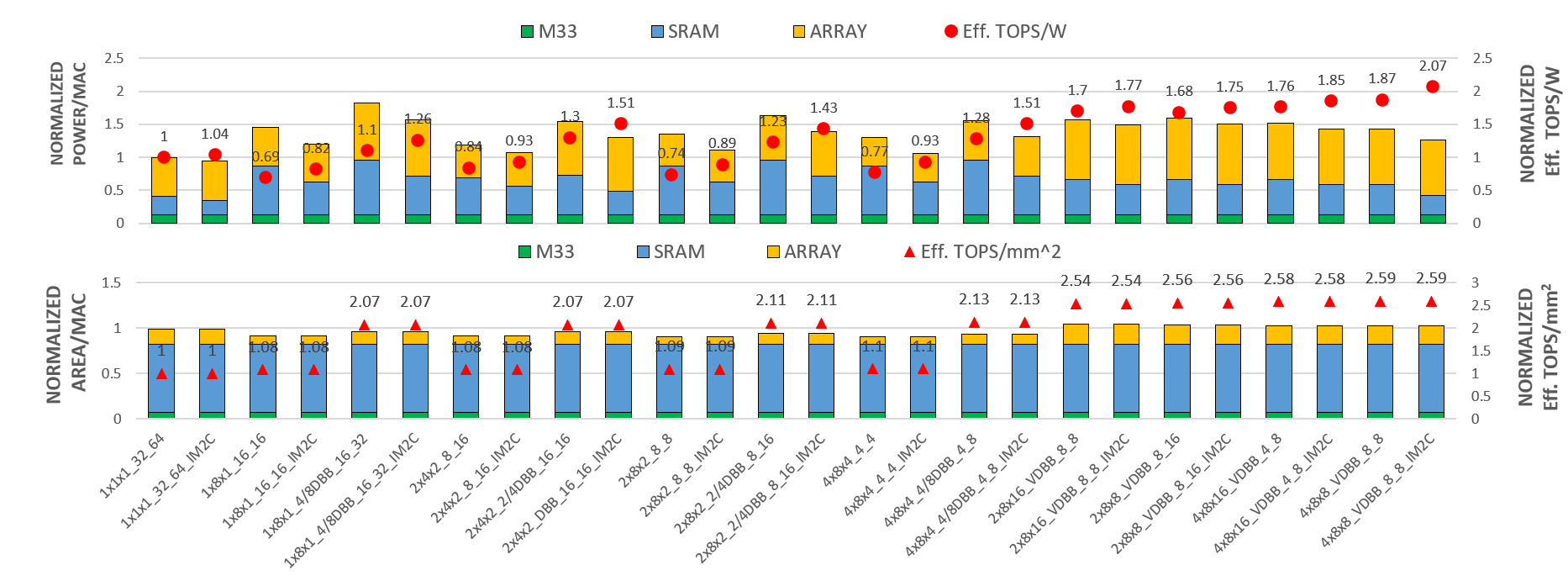}
    \caption{Power and area breakdown of iso-throughput designs at 4 TOPS peak, with 3/8 DBB (62.5\% sparsity) weights and 50\% random sparse activations.
    Normalized to a conventional 1\x1\x1 TPU-like systolic array baseline.
    }
    \label{fig:ppa-4-tops}
\end{figure*}

\section{Methodology} 
\label{sec:method}

\subsection{DBB CNN Training}
\label{sec:method:training}

Models must be specially trained to exploit DBB.
To demonstrate this, we trained five popular benchmark CNN models with INT8 DBB encoding on the standard image classification datasets MNIST~\cite{MNIST}, CIFAR-10~\cite{CIFAR10} and ImageNet~\cite{Imagenet}.
Our DBB training procedure consists of three phases.
Firstly, we start with pre-trained models,
although these can alternatively be trained from scratch using published recipes.
Secondly, we apply magnitude based DBB-aware weight pruning, which is similar to random pruning~\cite{zhu2017prune}, except that it operates within the domain of each DBB block.
This step progressively prunes small-magnitude weights within each DBB block for about 20 epochs, until the desired block sparsity constraint is met. 
As is conventional, the first convolution layer e.g. in MobileNet-V1, was not pruned, which has negligible impact on model size.
Finally, the model is fine tuned with 8-bit quantization for both DBB weights and activations, using the straight-through estimator (STE) approach which guarantees the FP value zero is converted precisely to integer value zero, for around 30 epochs.
We used TensorFlow 1.8.0 on an Intel Xeon server running Ubuntu Linux OS, with an Nvidia V100 GPU.

\subsection{RTL Generator}

The proposed accelerator (Section~\ref{sec:npu} encompasses a scalable family of configurations.
It is also regular and amenable to hierarchical implementation and validation.
We implemented a parameterized Python RTL generator to rapidly and precisely explore the full design space.
This generator produces synthesizable Verilog RTL for the accelerator, along with a testbench suite. 
Designs can be generated with arbitrary dimensions of A\x B\x C\_ M\x N, along with optional support for DBB and  VDBB sparsity, activation clock gating options etc. 
Each design is automatically validated in Synopsys VCS using the generated testbench, which can execute inference on a CNN model while logging value change dump (VCD) switching activity traces for each design on a given CNN.

\subsection{Physical Design and Evaluation}

The generated RTL was implemented in TSMC 16nm FinFET technology to evaluate circuit area, power dissipation and clock frequency.
We also implemented a design in TSMC 65nm LP bulk CMOS to allow fair comparison with results reported in the older technology.
The EDA tool flow used consists of Synopsys and Cadence tools, which we use with the TSMC PDK and Arm multi-Vt standard cell libraries.
The single-ported SRAM instances were carefully chosen from the options available in an Arm SRAM compiler.
The design was constrained for a 1GHz clock period at the slow corner, with multiple process, voltage and temperature corners used for setup and hold timing.
The 65nm design achieved 500MHz at the slow corner.
Power analysis was performed at the typical corner, using Synopsys PrimeTimePX, with the parasitic-annotated netlist and switching activity from VCD simulation traces.


While architecting and analyzing the performance of random-sparse accelerators~\cite{scnn} is challenging due to the potentially widely varying sparsity patterns, DBB sparse models have fixed sparsity and easily predictable runtime.
We evaluated each design using RTL simulation in Synopsys VCS running our DBB INT8 CNN models (Table ~\ref{tab:dbb-training}).
This generates accurate performance (throughput) metrics which vary depending on the dimensions of the weight and activation matrices in each layer.
For power consumption analysis, we capture VCD traces in RTL simulation from representative layers of ResNet50, which is then input to PrimeTimePX.

\section{Evaluation Results}
\label{sec:results}

In this section, we review the implementation results of proposed accelerators and compare with previous publications.

\begin{figure}[!t]
\centering
\includegraphics[width=0.95\columnwidth]{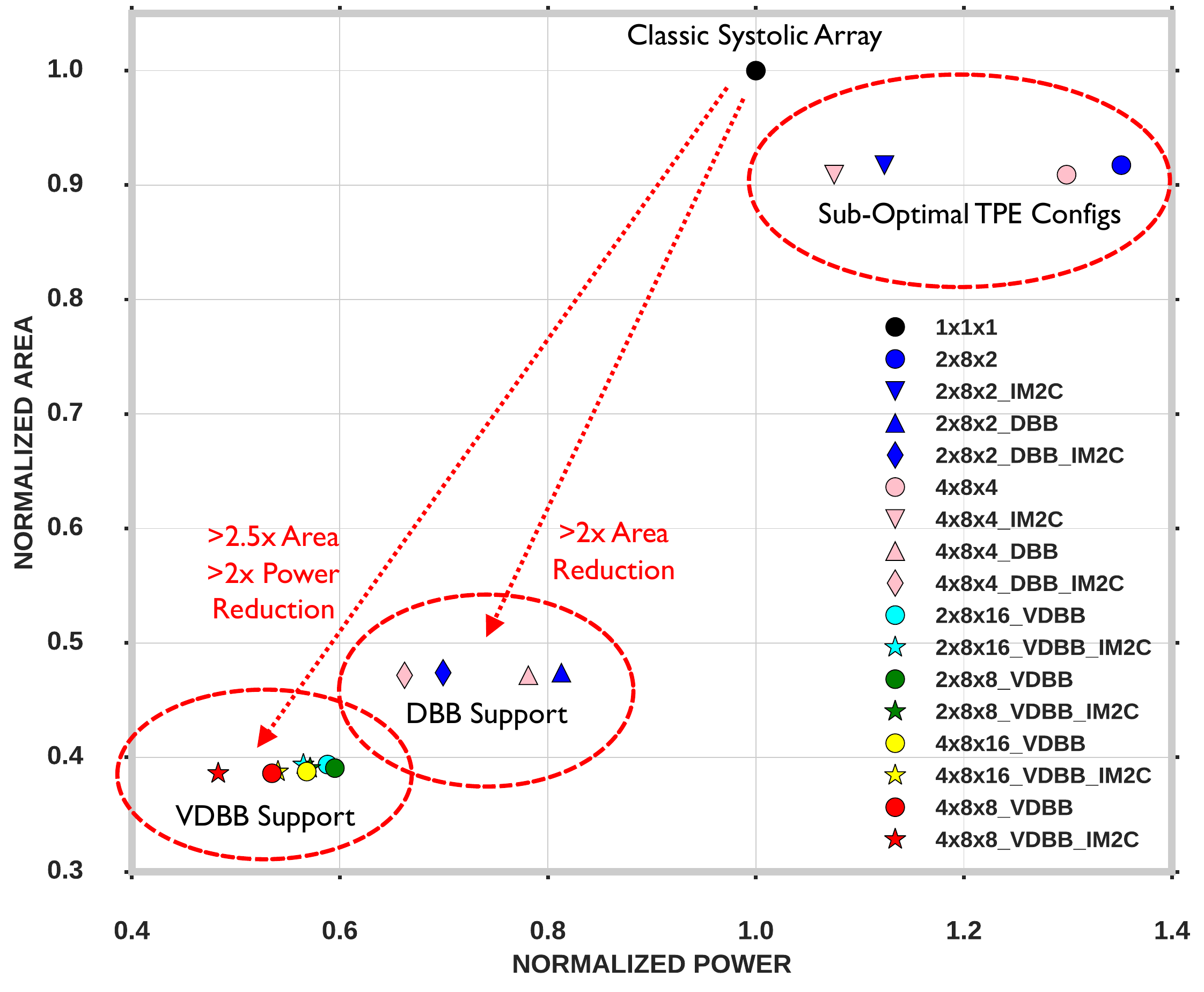}
\label{fig:power_area_4_tops}
\caption{
Effective power and area design space of the proposed accelerators, normalized to the 1\x1\x1 systolic array baseline.
The design space includes options for array configuration, hardware IM2COL, 3/8 DBB, and VDBB.
All points have 4 TOPS of nominal datapath performance, and include typical 50\% random activation sparsity.
}
\label{fig:power-area}
\end{figure}

\subsection{Design Space}
\label{sec:design_space_exploaration}

\begin{figure*}[!t]
\centering
\includegraphics[width=1\textwidth]{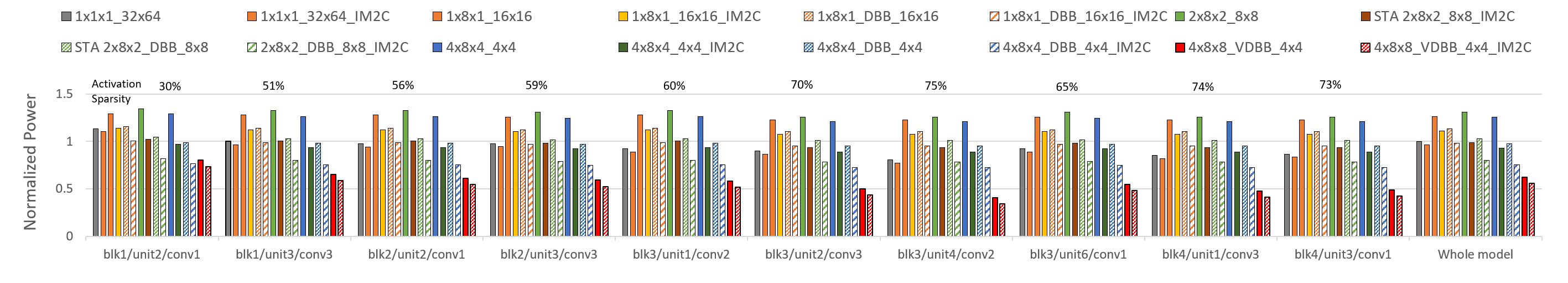}
\caption{4 TOPS nominal designs: power for individual layers and the whole model of INT8 DBB ResNet50\_V1. 
Normalized to 1\x1\x1 with 50\% average activation sparsity, closest to the  $\textit{blk1/unit3/conv3}$ layer in this ResNet50 example.
}
\label{fig:resnet_v1_layers}
\end{figure*}

The proposed microarchitecture described in Section~\ref{sec:npu}, has a number of parameters to be optimized, with some trade-offs (Section~\ref{sec:npu}).
We consider the design space for a typical 4 TOPS mobile CNN accelerator implementation.
Area and power consumption metrics are shown in Fig.~\ref{fig:ppa-4-tops}, where each design has equivalent peak throughput.
Each design point shown is described by a string which includes the array dimensions, the optional hardware IM2COL unit (denoted ``IM2C''), optional fixed DBB (``DBB''), and optional variable DBB (``VDBB'').
Not all combinations of these parameters are valid.
All the designs are configured to have the same peak throughput of 4 TOPS, and normalized to the baseline of conventional TPU-like systolic array, which we refer to as 1\x1\x1\_32\x64.
The TOPS/W and TOPS/mm$^2$ results assume a fairly typical 3/8 DBB weight sparsity and 50\% activation sparsity.
The best design is 4\x8\x8\_4\x8\_VDBB\_IM2C, which is summarized in full detail in Table~\ref{tab:best_design}.
This strong improvement is due to the significant advantages of exploiting sparsity and reuse in the microarchitecture (Table~\ref{tab:tpe}).

Fig.~\ref{fig:power-area} further illustrates the improvement from the optimizations, as it shows the design space of area and power, again normalized to the systolic array baseline.
This illustrates three distinct groupings of designs.
The first in the top right of the design space shows a cluster of sub-optimal array configurations (i.e. 2\x8\x2 and 4\x8\x4), without sparsity support.
The second group is in the bottom right, featuring designs with fixed DBB sparsity support, which achieve more than 2\x area reduction compared to the baseline.
Finally, the third group in the far bottom left are pareto-optimal VDBB designs, which benefit the most from IM2COL.
These designs improve the area by $>$2.5$\times$ and the power by $>$2$\times$.
The best design is summarized in detail in Table~\ref{tab:best_design}, showing power and area breakdown for each of the major components.

Unlike with random sparse weight accelerators~\cite{scnn}, the power consumption of proposed microarch. with DBB weights is fairly constant.
However, as we exploit random activation sparsity, the power varies from layer to layer on a real world model.
To illustrate this, Fig.~\ref{fig:resnet_v1_layers} shows normalized power for the popular ResNet-50-v1 model. 
Twelve designs  designs are shown, which are representative of the whole design space, all with 4 TOPS peak throughput.
All our accelerators take advantage of activation sparsity using a simple clock gating scheme.
The average activation sparsity percentage is annotated above each bar group.

Over the whole model, the 4\x8\x8\_VDBB\_IM2C design achieves a 44.6\% power reduction over the baseline.
Fixed DBB design 4\x8\x4\_DBB\_IM2C gives a 24.9\% power reduction.
Lastly, the variation over individual layers was mainly impacted by changes in activation sparsity.

\subsection{Variable DBB Sparsity}

\begin{figure}[!t]
    \centering
    \subfloat[Throughput vs. Model Sparsity]{
        \centering
        \includegraphics[width=1\columnwidth]{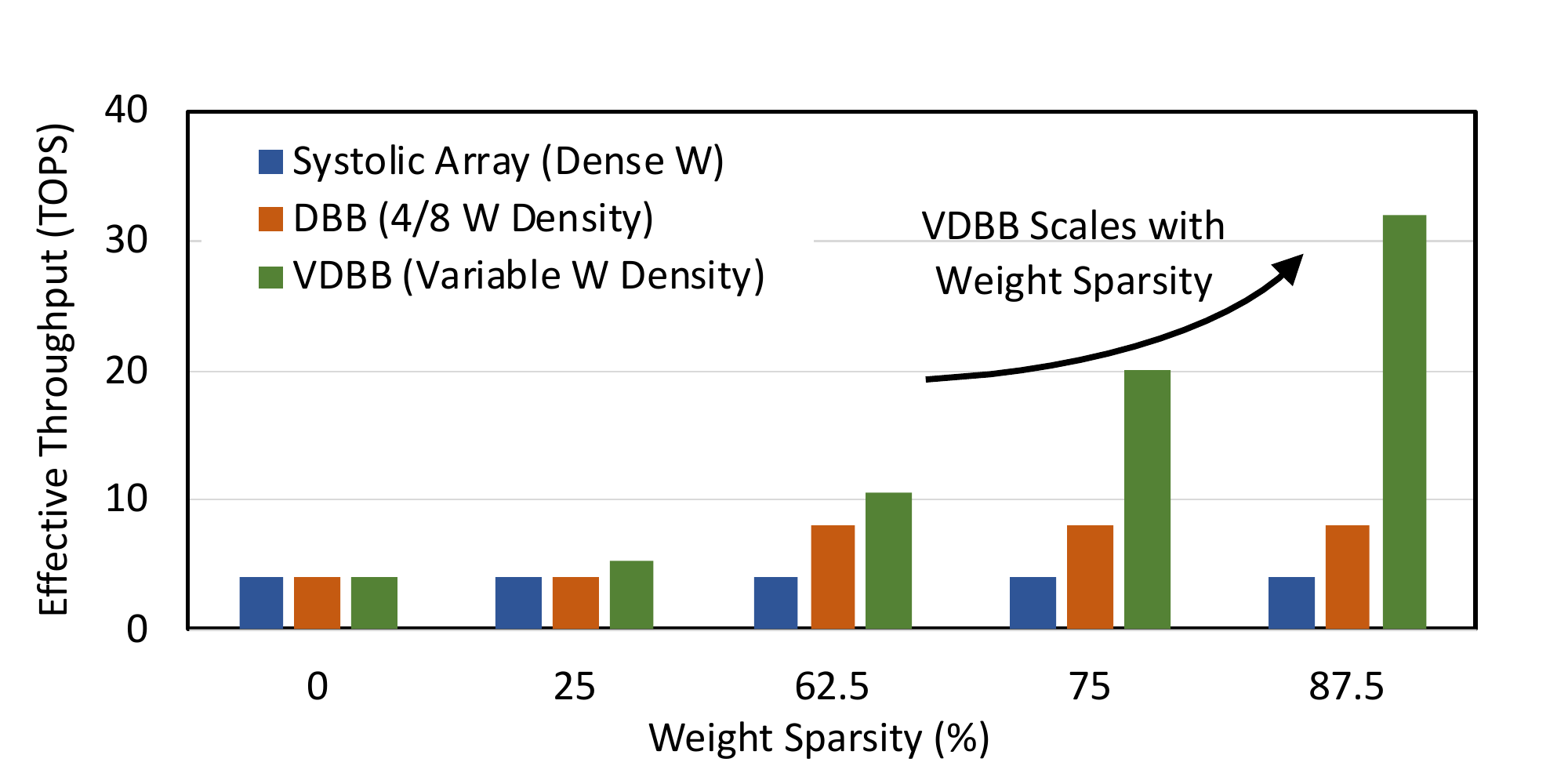}
        \label{fig:sparsity:throughput}
    }
    \vspace{0pt}
    \subfloat[Energy Efficiency vs. Model Sparsity]{
        \centering
        \includegraphics[width=1\columnwidth]{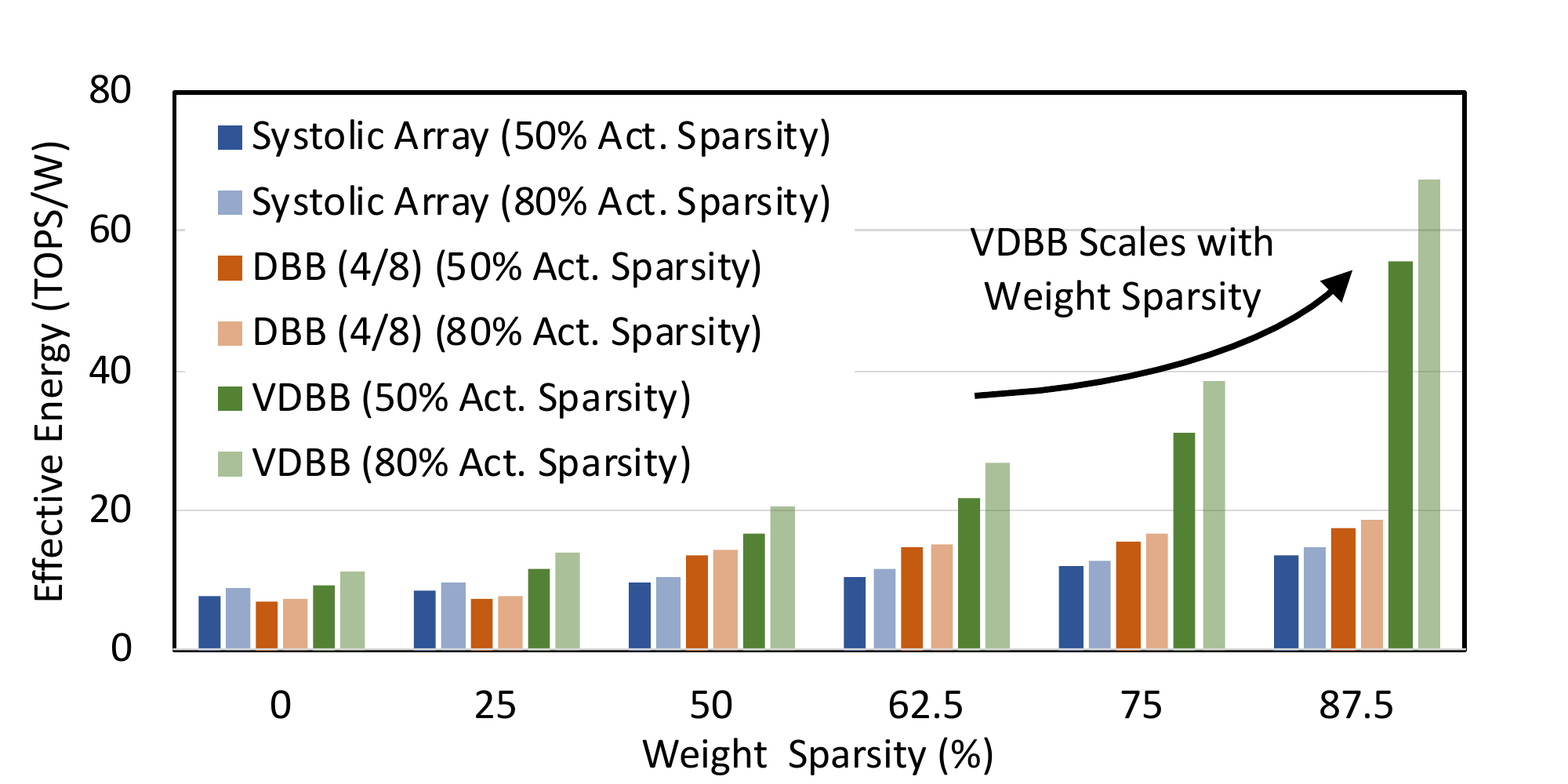}
        \label{fig:sparsity:energy}
    }
    \hspace{1pt}
\caption{
Scaling of (a) effective throughput and (b) effective energy, with weight sparsity for baseline systolic array with clock gating (1\x1\x1\_32\x64), DBB (4\x8\x4\_4\x8), and VDBB (4\x8\x4\_8\x8).
Energy efficiency increases with activation sparsity (50\% and 80\% shown), but does not impact throughput.
VDBB effectively exploits increasing weight sparsity, and outperforms both an optimized baseline (Systolic array with CG) and a fixed DBB implementation optimized for 4/8 density.
} 
\label{fig:sparsity}
\end{figure}

The key advantage of our proposal is that we support structured sparsity with a variable sparsity rate (Section~\ref{sec:background}), where as in the previous work this is fixed: $6/8$ in Kang~\cite{kang-tcsvt19} and $2/4$ in the Nvidia A100~\cite{nv-a100-datasheet}.
Fig~\ref{fig:sparsity} illustrates this point in terms of effective throughput and energy efficiency over the full range of weight sparsity levels available with a block size of 8.
Three designs are shown.
The first is the baseline systolic array (1\x1\x1 array), optimized with hardware IM2COL and activation sparsity CG.
The second is the design (Table~\ref{tab:best_design}) with fixed 4/8 DBB.
The third one is proposed VDBB architecture with variable DBB.
Results are given with typical 50\% and 80\% activation sparsity.
The systolic array baseline shows no throughput improvement as model sparsity increases, but does show some energy improvement due to less switching activity in the circuits.
The fixed DBB accelerator shows a step improvement in throughput at and above 50\% sparsity, which corresponds with the fixed 4/8 DBB design point.
The energy is similar, but again shows a little more improvement at high sparsity due to reduced switching.
Finally, the variable DBB (VDBB) accelerator successfully demonstrates throughput and energy which scale very strongly with the model sparsity.
Around 50\% model sparsity, VDBB is slightly superior to fixed DBB, but above 50\% the benefit is very large indeed.
Models optimized for very high weight sparsity of 87.5\% can achieve as much as 30 TOPS effective throughput and 56 TOPS/W energy efficiency.
This confirms the advantage of the time unrolled VDBB architecture.

\subsection{Comparison with Prior Work}

\begin{table}[t]
\centering
\begin{tabular}{l l l }
\toprule
Component                       &  Power, $mW$ (\%)             & Area, $mm^2$ (\%) \\
\midrule    
Systolic Tensor Array              &  318 (65.2\%)                & 0.732 (20.0\%)  \\ 
Weight SRAM (512KB)                &  78.5 (16.1\%)                  & 0.54 (14.4\%)   \\ 
Activation SRAM (2MB)              &  31.0 (6.4\%), 93.0$^\dagger$  & 2.16 (57.8\%)  \\
Cortex-M33 MCU~\cite{arm_m33} \x4. &  50.5 (10.2\%)               & 0.30 (8.0\%)  \\ 
IM2COL Unit                        &  10.0 (2\%)              & 0.01 (0.26\%) \\
\midrule
Total                              &  487.5 (100\%), 539.5$^\dagger$    &  3.74 (100\%)  \\
\bottomrule
\end{tabular}
\vspace{1pt}
\\ $^\dagger$ IM2COL disabled
\caption{
Summary of the pareto-optimal VDBB design: 4$\times$8$\times$8\_4$\times$8\_VDBB\_IM2COL, with nominal 4 TOPS.
Power reported with nominal 3/8 (62.5\%) DBB model sparsity and 50\% random sparse activations, achieving an effective 21.9 TOPS/W and 2.85 TOPS/mm$^2$.    
}
\label{tab:best_design}
\vspace{20pt}
\end{table}

\begin{table*}[t]
\centering
\begin{tabular}{l c c c c c c c c}
\toprule
                                & Technology         & SRAM          & Freq. & Throughput    & Energy Efficiency$^1$     & Area Efficiency$^1$   & Weight    & Activation \\
                                &               & A / W         & (GHz)   & (TOPS)      & (TOPS/W)                  & (TOPS/mm$^2$)         & Sparsity  & Sparsity   \\
\midrule 
Ours                    & 16nm          & 2MB / 512KB   & 1.0      & 4             & 55.7                      & 8.52                   &  87.5\% VDBB        & 50\% CG         \\
                                & 16nm          & 2MB / 512KB   & 1.0      & 4             & 31.3                      & 4.29                   &  75\% VDBB        & 50\% CG         \\
                               & 16nm          & 2MB / 512KB   & 1.0      & 4             & 21.9                      & 2.85                   &  62.5\% VDBB        & 50\% CG       \\
                               & 16nm          & 2MB / 512KB   & 1.0        & 4             & 16.8                      & 2.13                   &  50\% VDBB        & 50\% CG         \\
SMT-SA$^2$~\cite{smt-sa}        & 16nm          & 2MB / 512KB   & 1.0      & 4              & 7.4                      & 1.13                    &  62.5\% Random   &  50\% CG                       \\
Laconic~\cite{laconic_isca19}   & 15nm          & 2MB / 512KB   & 1.0      & --              & 1.997                     & --                    &  Bit-wise &    Bit-wise               \\
SCNN$^3$~\cite{laconic_isca19}  & 16nm      & 1.2MB / --          & 1.0    & 2              & 0.79                      & 0.7                    & Random            &      --                 \\
\midrule
Ours                      & 65nm       & 2MB / 512KB   & 0.5      &  1             & 2.80                    & 0.26                & 75\% VDBB          &   50\% CG                    \\
                           & 65nm       & 2MB / 512KB   & 0.5      &    1          & 1.95                    & 0.17                & 62.5\% VDBB          &   50\% CG                    \\
Kang et al.~\cite{kang-tcsvt19} & 65nm          & 58KB        &   1.0       &    0.5          & 1.65                      & 1.01                  & 75\% DBB          &         --              \\  
Laconic~\cite{laconic_isca19}   & 65nm          & 2MB / 512KB.  &    1.0      &        --       & 0.81                      & --                    & Bit-wise          &    Bit-wise                   \\
Eyeriss v2~\cite{eyeriss_v2}    & 65nm       & 246KB         & 0.2    & 0.40              & 0.96                      & 0.07/2.7M gates                    & Random            &    Random                   \\

\bottomrule
\end{tabular}
\\
\vspace{3pt}
$^1$ Effective operations.
$^2$ Our re-implementation with INT8 operands in 16nm.
$^3$ 24-bit accumulators rather than 32-bit.
\caption{Comparison with published sparse INT8 CNN accelerators in 16nm and 65nm technology.
Published metrics for these works varies wildly; however, even at modest 50\% weight and activation sparsity, Our VDBB design achieves 16.8 TOPS/W in 16nm, which far exceeds previously reported results and offers strong throughput and energy scaling with weight sparsity.
}
\label{tab:accelerator}
\vspace{10pt}
\end{table*}

This work has demonstrated a time unrolled variable DBB scheme (Section~\ref{sec:vdbb}) implemented in a reuse optimized accelerator (Section~\ref{sec:npu}).
Here, we illustrate the benefit of these two contributions by comparing our work with previously published INT8 CNN inference accelerators in Table~\ref{tab:accelerator}.
The proposed designs shown include the nominal 4 TOPS design in Table~\ref{tab:best_design} at multiple model sparsity points, as well as a 65nm version of the same design to allow a broader comparison with results in the older technology.

We first evaluate our results specifically against the state-of-the-art sparse CNN accelerator work, Laconic~\cite{laconic_isca19}.
This paper includes a thorough survey against the latest work, including: DaDianNao++~\cite{dadiannao2014micro}, Eyeriss~\cite{eyerissIsca}, SCNN~\cite{scnn}, Pragmatic~\cite{pragmatic_micro17}, and BitFusion~\cite{bitfusion_isca18}, which convincingly demonstrates it is the current state-of-the-art.
Therefore, this is a useful comparison point at the same INT8 precision, 1 GHz clock frequency and comparable 15nm technology.
The energy efficiency result reported for Laconic is just under 2 TOPS/W energy efficiency.
Our nominal 4 TOPS VDBB design (Table~\ref{tab:best_design}) achieves 16.8 TOPS/W at 50\% model sparsity, which is more than 8$\times$ higher.


The only other DBB accelerator is Kang~\cite{kang-tcsvt19}, which uses a fixed $2/8$ DBB implemented in a dot-product microarchitecture, and reports 1.65 TOPS/W in 65nm technology for 75\% sparse DBB.
Our design implements variable DBB in a reuse optimized accelerator and in 65nm achieves 2.8 TOPS/W at the same 75\% model sparsity, which is 70\% higher.
We also note that such a high fixed DBB sparsity would not be practical for compact models like MobileNet, ResNet on the ImageNet dataset, based on our results (Table~\ref{tab:dbb-training}).

The only other sparse systolic array design we are aware of is SMT-SA~\cite{smt-sa}.
To compare against this work, which was reported in 45nm, we implemented the same design ourselves, which achieves 7.4 TOPS/W compared to the proposed design at 21.9 for the same sparsity.
This is largely due to the cost of the FIFOs required in the array, and the advantages of DBB vs random weight sparsity (Section~\ref{sec:background}).

Finally, SCNN~\cite{scnn} and Eyeriss v2~\cite{eyeriss_v2} are also included in the comparison (Table~\ref{tab:accelerator}).
In summary, we found that our work outperforms: 1) Laconic~\cite{laconic_isca19}, the latest sparse accelerator, 2) SMT-SA~\cite{smt-sa}, the only other sparse systolic array, and 3) Kang~\cite{kang-tcsvt19}, the only other DBB accelerator.



\section{Related Work}
\label{sec:related}

\paragraph{Clock Gating Random Sparsity}
A simple and effective approach to exploiting random sparsity is to clock-gate (CG) to save power when one or more zero operands are encountered~\cite{eyerissIsca,minerva}.
However, CG schemes generally result in low utilization with no area or throughput improvement.
We apply this CG for activation sparsity (which is not amenable to DBB).

\paragraph{Random Sparsity}
For RNN acceleration, EIE~\cite{eie} implements a fine-grained random sparse CSR-encoded INT16 matrix-vector accelerator for dense layers, and ESE~\cite{ese_fpga17} extends this to LSTMs.
MASR~\cite{masr_pact19} also exploits random sparsity, but uses a bitmask encoding.
PermDNN~\cite{deng-micro18} targets sparse dense layers using permuted diagonal matrices.
A number of papers target random sparse matrix multiplication for very sparse data, such as 
Outer Space~\cite{Outer-space} which uses an outer product scheme, and SpArch~\cite{zhang2020sparch}, which further optimizes for locality.
More specific to the lower sparsity of CNNs, Cnvlutin~\cite{albericio2016isca} demonstrates skipping compute for zero activations discovered at runtime, without explicit indexes.
SCNN~\cite{scnn} implements a fully CSR-indexed sparse CNN accelerator using an outer product to exploit sparse weights and activations.
FixyNN~\cite{fixy2019sysml} demonstrates a fixed-weight accelerator that can very efficiently exploit random sparsity.
We focus on CNN structured sparsity, but compare with SCNN and Laconic (Table~\ref{tab:accelerator}).

\paragraph{Structured Sparsity} 
Cambricon-S proposes a conventional block sparse accelerator~\cite{zhou-micro18}.
A DBB accelerator described by Kang~\cite{kang-tcsvt19} implements a fixed weight sparsity of 75\%.
The accelerator design is also based on a dot product microarchitecture with limited data reuse.
The hardware implementation of the Nvidia structured sparsity scheme~\cite{nv-a100-datasheet} are unknown, but is fixed at $2/4$ (50\%) sparsity.
From our pruning experiments (Table~\ref{tab:dbb-training}), $2/8$ (75\%) is quite aggressive, but $2/4$ is probably more universally useful.
Nonetheless, in both cases, the deployed benefit is limited due to the fixed-sparsity ratio.
In contrast, our VDBB proposal demonstrates variable-sparsity DBB using time unrolling in a reuse optimized accelerator.

\paragraph{Sparsity in Systolic Arrays}
Most sparse CNN accelerators are based on dot-product designs reminiscent of DaDianNao~\cite{dadiannao2014micro}, which typically have lower data reuse compared to systolic arrays (SAs) like the Google TPU~\cite{tpu-isca}.
SMT-SA~\cite{smt-sa} is a sparse SA, which foccusses on random sparsity (instead of DBB).
Kung et al.~\cite{kung_2018} demonstrated a preprocessing step of column combining of sparse weight matrices, before processing on a dense SA architecture.
We implemented an INT8 version of SMT-SA to compare against (Table~\ref{tab:accelerator}), and found that the DBB is much more efficient than the random sparsity of SMT-SA, which requires FIFOs in the array.


\section{Conclusion}
\label{sec:conclusion}

Structured model sparsity is a powerful optimization to enable improved throughput and energy efficiency in CNN hardware accelerators, without the overheads and unpredictable load balancing of random weight sparsity.
However, unlike with random sparsity, previous demonstrations of block sparsity employ a fixed target sparsity ratio.
Unfortunately, this is a serious impediment to broad deployment, because real world CNNs typically exhibit a wide range of weight sparsity ratios.
With a fixed sparsity, any models that do not achieve this threshold must fall back to dense operation with no speedup.
At the same time, aggressively optimized models that exceed the threshold also do not see any benefit.

In this paper, we introduce a variable density bound block (VDBB) technique, which uses a time unrolled architecture to implement a weight sparsity scheme that supports all possible block sparsity levels.
This enables hardware benefits from model sparsity across the whole spectrum of models in use today.
We implement VDBB in a reuse optimized accelerator microarchitecture, featuring a systolic tensor array (STA) composed of more complex PEs called tensor PEs (TPEs) which increase operand and accumulator reuse.
We also describe a hardware delayed IM2COL unit that achieves a 3\x activation SRAM bandwidth magnification effect to reduce SRAM power consumption.
The reuse optimized accelerator design introduces a number of interdependent parameters, resulting in a non-trivial design space, which we evaluate in 16nm process technology.
The optimal design scales strongly in throughput and energy as a function of model sparsity, from 16.8 TOPS/W at 50\% sparsity up to 55.7 TOPS/W at 87.5\%.
The advantage of both the VDBB compression and the reuse optimizations is apparent in these results, which outperform previous sparse CNN accelerator designs reported.




%% file: main.bbl
\begin{thebibliography}{10}
\providecommand{\url}[1]{#1}
\csname url@samestyle\endcsname
\providecommand{\newblock}{\relax}
\providecommand{\bibinfo}[2]{#2}
\providecommand{\BIBentrySTDinterwordspacing}{\spaceskip=0pt\relax}
\providecommand{\BIBentryALTinterwordstretchfactor}{4}
\providecommand{\BIBentryALTinterwordspacing}{\spaceskip=\fontdimen2\font plus
\BIBentryALTinterwordstretchfactor\fontdimen3\font minus
  \fontdimen4\font\relax}
\providecommand{\BIBforeignlanguage}[2]{{%
\expandafter\ifx\csname l@#1\endcsname\relax
\typeout{** WARNING: IEEEtranS.bst: No hyphenation pattern has been}%
\typeout{** loaded for the language `#1'. Using the pattern for}%
\typeout{** the default language instead.}%
\else
\language=\csname l@#1\endcsname
\fi
#2}}
\providecommand{\BIBdecl}{\relax}
\BIBdecl

\bibitem{arm_m33}
\BIBentryALTinterwordspacing
{Arm Cortex-M33}. [Online]. Available:
  \url{https://developer.arm.com/ip-products/processors/cortex-m/cortex-m33}
\BIBentrySTDinterwordspacing

\bibitem{arm_v8m}
\BIBentryALTinterwordspacing
{ARMv8-M Architecture Reference Manual}. [Online]. Available:
  \url{https://static.docs.arm.com/ddi0553/a/DDI0553A_e_armv8m_arm.pdf}
\BIBentrySTDinterwordspacing

\bibitem{mobilenetv1_model}
\BIBentryALTinterwordspacing
{MobileNet v1 TensorFlow Model}. [Online]. Available:
  \url{http://download.tensorflow.org/models/mobilenet_v1_2018_08_02/mobilenet_v1_1.0_224.tgz}
\BIBentrySTDinterwordspacing

\bibitem{pragmatic_micro17}
\BIBentryALTinterwordspacing
J.~Albericio, A.~Delm\'{a}s, P.~Judd, S.~Sharify, G.~O’Leary, R.~Genov, and
  A.~Moshovos, ``Bit-pragmatic deep neural network computing,'' in
  \emph{Proceedings of the 50th Annual IEEE/ACM International Symposium on
  Microarchitecture}, ser. MICRO-50 ’17.\hskip 1em plus 0.5em minus
  0.4em\relax New York, NY, USA: Association for Computing Machinery, 2017, p.
  382–394. [Online]. Available: \url{https://doi.org/10.1145/3123939.3123982}
\BIBentrySTDinterwordspacing

\bibitem{albericio2016isca}
\BIBentryALTinterwordspacing
J.~Albericio, P.~Judd, T.~Hetherington, T.~Aamodt, N.~E. Jerger, and
  A.~Moshovos, ``Cnvlutin: Ineffectual-neuron-free deep neural network
  computing,'' in \emph{Proceedings of the 43rd International Symposium on
  Computer Architecture}, ser. ISCA '16.\hskip 1em plus 0.5em minus 0.4em\relax
  Piscataway, NJ, USA: IEEE Press, 2016, pp. 1--13. [Online]. Available:
  \url{https://doi.org/10.1109/ISCA.2016.11}
\BIBentrySTDinterwordspacing

\bibitem{eyeriss_v2}
Y.~{Chen}, T.~{Yang}, J.~{Emer}, and V.~{Sze}, ``Eyeriss v2: A flexible
  accelerator for emerging deep neural networks on mobile devices,'' \emph{IEEE
  Journal on Emerging and Selected Topics in Circuits and Systems}, vol.~9,
  no.~2, pp. 292--308, 2019.

\bibitem{eyerissIsca}
\BIBentryALTinterwordspacing
Y.-H. Chen, J.~Emer, and V.~Sze, ``Eyeriss: A spatial architecture for
  energy-efficient dataflow for convolutional neural networks,'' in
  \emph{Proceedings of the 43rd International Symposium on Computer
  Architecture}.\hskip 1em plus 0.5em minus 0.4em\relax Piscataway, NJ, USA:
  IEEE Press, 2016, pp. 367--379. [Online]. Available:
  \url{https://doi.org/10.1109/ISCA.2016.40}
\BIBentrySTDinterwordspacing

\bibitem{dadiannao2014micro}
\BIBentryALTinterwordspacing
Y.~Chen, T.~Luo, S.~Liu, S.~Zhang, L.~He, J.~Wang, L.~Li, T.~Chen, Z.~Xu,
  N.~Sun, and O.~Temam, ``Dadiannao: A machine-learning supercomputer,'' in
  \emph{Proceedings of the 47th Annual IEEE/ACM International Symposium on
  Microarchitecture}, ser. MICRO-47.\hskip 1em plus 0.5em minus 0.4em\relax
  Washington, DC, USA: IEEE Computer Society, 2014, pp. 609--622. [Online].
  Available:
  \url{http://dx.doi.org.ezp-prod1.hul.harvard.edu/10.1109/MICRO.2014.58}
\BIBentrySTDinterwordspacing

\bibitem{xception}
\BIBentryALTinterwordspacing
F.~Chollet, ``Xception: Deep learning with depthwise separable convolutions,''
  \emph{CoRR}, vol. abs/1610.02357, 2016. [Online]. Available:
  \url{http://arxiv.org/abs/1610.02357}
\BIBentrySTDinterwordspacing

\bibitem{deng-micro18}
C.~{Deng}, S.~{Liao}, Y.~{Xie}, K.~K. {Parhi}, X.~{Qian}, and B.~{Yuan},
  ``Permdnn: Efficient compressed dnn architecture with permuted diagonal
  matrices,'' in \emph{2018 51st Annual IEEE/ACM International Symposium on
  Microarchitecture (MICRO)}, 2018, pp. 189--202.

\bibitem{Imagenet}
J.~Deng, W.~Dong, R.~Socher, L.-J. Li, K.~Li, and L.~Fei-Fei, ``{ImageNet: A
  Large-Scale Hierarchical Image Database},'' in \emph{CVPR09}, 2009.

\bibitem{fedorov2019sparse}
I.~Fedorov, R.~P. Adams, M.~Mattina, and P.~N. Whatmough, ``{SpArSe: Sparse
  architecture search for CNNs on resource-constrained microcontrollers},'' in
  \emph{Advances in Neural Information Processing Systems (NeurIPS)}, 2019, pp.
  4978--4990.

\bibitem{tinylstm}
I.~Fedorov, M.~Stamenovic, C.~Jenson, L.-C. Yang, A.~Mandell, Y.~Gan,
  M.~Mattina, and P.~N. Whatmough, ``{ TinyLSTMs: Efficient Neural Speech
  Enhancement for Hearing Aids },'' in \emph{Conference of the International
  Speech Communication Association (INTERSPEECH)}, 2020.

\bibitem{asv}
\BIBentryALTinterwordspacing
Y.~Feng, P.~Whatmough, and Y.~Zhu, ``{ASV: Accelerated Stereo Vision System},''
  in \emph{Proceedings of the 52nd Annual IEEE/ACM International Symposium on
  Microarchitecture}, ser. MICRO '52.\hskip 1em plus 0.5em minus 0.4em\relax
  New York, NY, USA: Association for Computing Machinery, 2019, p. 643–656.
  [Online]. Available: \url{https://doi.org/10.1145/3352460.3358253}
\BIBentrySTDinterwordspacing

\bibitem{masr_pact19}
\BIBentryALTinterwordspacing
U.~Gupta, B.~Reagen, L.~Pentecost, M.~Donato, T.~Tambe, A.~M. Rush, G.~Wei, and
  D.~Brooks, ``{MASR:} {A} modular accelerator for sparse rnns,'' in \emph{28th
  International Conference on Parallel Architectures and Compilation
  Techniques, {PACT} 2019, Seattle, WA, USA, September 23-26, 2019}.\hskip 1em
  plus 0.5em minus 0.4em\relax {IEEE}, 2019, pp. 1--14. [Online]. Available:
  \url{https://doi.org/10.1109/PACT.2019.00009}
\BIBentrySTDinterwordspacing

\bibitem{eie}
S.~Han \emph{et~al.}, ``{EIE}: Efficient inference engine on compressed deep
  neural network,'' in \emph{Proceedings of the 43rd Int. Symp. on Computer
  Architecture (ISCA)}, 2016.

\bibitem{ese_fpga17}
\BIBentryALTinterwordspacing
S.~Han, J.~Kang, H.~Mao, Y.~Hu, X.~Li, Y.~Li, D.~Xie, H.~Luo, S.~Yao, Y.~Wang,
  H.~Yang, and W.~B.~J. Dally, ``Ese: Efficient speech recognition engine with
  sparse lstm on fpga,'' in \emph{Proceedings of the 2017 ACM/SIGDA
  International Symposium on Field-Programmable Gate Arrays}, ser. FPGA
  ’17.\hskip 1em plus 0.5em minus 0.4em\relax New York, NY, USA: Association
  for Computing Machinery, 2017, p. 75–84. [Online]. Available:
  \url{https://doi.org/10.1145/3020078.3021745}
\BIBentrySTDinterwordspacing

\bibitem{deepcompression}
\BIBentryALTinterwordspacing
S.~Han, H.~Mao, and W.~J. Dally, ``Deep compression: Compressing deep neural
  network with pruning, trained quantization and huffman coding,'' \emph{CoRR},
  vol. abs/1510.00149, 2015. [Online]. Available:
  \url{http://arxiv.org/abs/1510.00149}
\BIBentrySTDinterwordspacing

\bibitem{hansen-icpr20}
P.~Hansen, A.~Vilkin, Y.~Khrustalev, J.~Imber, D.~Hanwell, M.~Mattina, and
  P.~N. Whatmough, ``{ ISP4ML: Understanding the Role of Image Signal
  Processing in Efficient Deep Learning Vision Systems },'' in
  \emph{International Conference on Pattern Recognition (ICPR)}, 2020.

\bibitem{tpu-isca}
\BIBentryALTinterwordspacing
N.~P. Jouppi, C.~Young, N.~Patil, D.~Patterson, G.~Agrawal, R.~Bajwa, S.~Bates,
  S.~Bhatia, N.~Boden, A.~Borchers, R.~Boyle, P.~Cantin, C.~Chao, C.~Clark,
  J.~Coriell, M.~Daley, M.~Dau, J.~Dean, B.~Gelb, T.~V. Ghaemmaghami,
  R.~Gottipati, W.~Gulland, R.~Hagmann, R.~C. Ho, D.~Hogberg, J.~Hu, R.~Hundt,
  D.~Hurt, J.~Ibarz, A.~Jaffey, A.~Jaworski, A.~Kaplan, H.~Khaitan, A.~Koch,
  N.~Kumar, S.~Lacy, J.~Laudon, J.~Law, D.~Le, C.~Leary, Z.~Liu, K.~Lucke,
  A.~Lundin, G.~MacKean, A.~Maggiore, M.~Mahony, K.~Miller, R.~Nagarajan,
  R.~Narayanaswami, R.~Ni, K.~Nix, T.~Norrie, M.~Omernick, N.~Penukonda,
  A.~Phelps, J.~Ross, A.~Salek, E.~Samadiani, C.~Severn, G.~Sizikov,
  M.~Snelham, J.~Souter, D.~Steinberg, A.~Swing, M.~Tan, G.~Thorson, B.~Tian,
  H.~Toma, E.~Tuttle, V.~Vasudevan, R.~Walter, W.~Wang, E.~Wilcox, and D.~H.
  Yoon, ``In-datacenter performance analysis of a tensor processing unit,''
  \emph{CoRR}, vol. abs/1704.04760, 2017. [Online]. Available:
  \url{http://arxiv.org/abs/1704.04760}
\BIBentrySTDinterwordspacing

\bibitem{kang-tcsvt19}
H.~{Kang}, ``Accelerator-aware pruning for convolutional neural networks,''
  \emph{IEEE Transactions on Circuits and Systems for Video Technology}, pp.
  1--1, 2019.

\bibitem{kodali-iccd17}
S.~{Kodali}, P.~{Hansen}, N.~{Mulholland}, P.~{Whatmough}, D.~{Brooks}, and
  G.~{Wei}, ``{Applications of Deep Neural Networks for Ultra Low Power IoT},''
  in \emph{2017 IEEE International Conference on Computer Design (ICCD)}, 2017,
  pp. 589--592.

\bibitem{CIFAR10}
A.~Krizhevsky and G.~Hinton, ``Learning multiple layers of features from tiny
  images,'' \emph{Master's thesis, Department of Computer Science, University
  of Toronto}, 2009.

\bibitem{kung_2018}
H.~Kung, B.~McDanel, and S.~Q. Zhang, ``Packing sparse convolutional neural
  networks for efficient systolic array implementations: Column combining under
  joint optimization,'' in \emph{24th Int. Conf. on Architectural Support for
  Programming Languages and Operating Systems (ASPLOS)}, 2019, pp. 821--834.

\bibitem{MNIST}
Y.~{Lecun}, L.~{Bottou}, Y.~{Bengio}, and P.~{Haffner}, ``Gradient-based
  learning applied to document recognition,'' \emph{Proceedings of the IEEE},
  vol.~86, no.~11, pp. 2278--2324, 1998.

\bibitem{li-dac19}
H.~{Li}, M.~{Bhargav}, P.~N. {Whatmough}, and H.~. {Philip Wong}, ``On-chip
  memory technology design space explorations for mobile deep neural network
  accelerators,'' in \emph{2019 56th ACM/IEEE Design Automation Conference
  (DAC)}, 2019, pp. 1--6.

\bibitem{liu-cal20}
Z.~{Liu}, P.~N. {Whatmough}, and M.~{Mattina}, ``{Systolic Tensor Array: An
  Efficient Structured-Sparse GEMM Accelerator for Mobile CNN Inference},''
  \emph{IEEE Computer Architecture Letters}, vol.~19, no.~1, pp. 34--37, 2020.

\bibitem{nv-a100-datasheet}
\BIBentryALTinterwordspacing
Nvidia. {A100 GPU Datasheet}. [Online]. Available:
  \url{https://www.nvidia.com/content/dam/en-zz/Solutions/Data-Center/a100/pdf/nvidia-a100-datasheet.pdf}
\BIBentrySTDinterwordspacing

\bibitem{Outer-space}
S.~{Pal} \emph{et~al.}, ``Outerspace: An outer product based sparse matrix
  multiplication accelerator,'' in \emph{Int. Symp. on High Performance
  Computer Architecture (HPCA)}, Feb 2018, pp. 724--736.

\bibitem{scnn}
A.~{Parashar} \emph{et~al.}, ``{SCNN: An accelerator for compressed-sparse
  convolutional neural networks},'' in \emph{2017 ACM/IEEE 44th Annual
  International Symposium on Computer Architecture (ISCA)}, June 2017, pp.
  27--40.

\bibitem{minerva}
B.~Reagen, P.~Whatmough, R.~Adolf, S.~Rama, H.~Lee, S.~K. Lee, J.~M.
  Hern\'andez-Lobato, G.-Y. Wei, and D.~Brooks, ``Minerva: Enabling low-power,
  highly-accurate deep neural network accelerators,'' in \emph{Proceedings of
  the 43rd International Symposium on Computer Architecture}, ser. ISCA, 2016.

\bibitem{laconic_isca19}
\BIBentryALTinterwordspacing
S.~Sharify, A.~D. Lascorz, M.~Mahmoud, M.~Nikolic, K.~Siu, D.~M. Stuart,
  Z.~Poulos, and A.~Moshovos, ``Laconic deep learning inference acceleration,''
  in \emph{Proceedings of the 46th International Symposium on Computer
  Architecture}, ser. ISCA ’19.\hskip 1em plus 0.5em minus 0.4em\relax New
  York, NY, USA: Association for Computing Machinery, 2019, p. 304–317.
  [Online]. Available: \url{https://doi.org/10.1145/3307650.3322255}
\BIBentrySTDinterwordspacing

\bibitem{bitfusion_isca18}
\BIBentryALTinterwordspacing
H.~Sharma, J.~Park, N.~Suda, L.~Lai, B.~Chau, V.~Chandra, and H.~Esmaeilzadeh,
  ``Bit fusion: Bit-level dynamically composable architecture for accelerating
  deep neural networks,'' in \emph{Proceedings of the 45th Annual International
  Symposium on Computer Architecture}, ser. ISCA ’18.\hskip 1em plus 0.5em
  minus 0.4em\relax IEEE Press, 2018, p. 764–775. [Online]. Available:
  \url{https://doi.org/10.1109/ISCA.2018.00069}
\BIBentrySTDinterwordspacing

\bibitem{smt-sa}
G.~Shomron, T.~Horowitz, and U.~Weiser, ``{SMT-SA}: {S}imultaneous
  multithreading in systolic arrays,'' \emph{IEEE Comput. Archit. Lett.},
  vol.~18, no.~2, pp. 99--102, Jul. 2019.

\bibitem{warden_gemm}
\BIBentryALTinterwordspacing
P.~Warden. {Why GEMM is at the heard of deep learning}. [Online]. Available:
  \url{https://petewarden.com/2015/04/20/why-gemm-is-at-the-heart-of-deep-learning/}
\BIBentrySTDinterwordspacing

\bibitem{smiv}
P.~N. {Whatmough}, S.~K. {Lee}, M.~{Donato}, H.~{Hsueh}, S.~{Xi}, U.~{Gupta},
  L.~{Pentecost}, G.~G. {Ko}, D.~{Brooks}, and G.~{Wei}, ``{A 16nm 25mm2 SoC
  with a 54.5x Flexibility-Efficiency Range from Dual-Core Arm Cortex-A53 to
  eFPGA and Cache-Coherent Accelerators},'' in \emph{2019 Symposium on VLSI
  Circuits}, 2019, pp. C34--C35.

\bibitem{fixy2019sysml}
P.~N. Whatmough, C.~Zhou, P.~Hansen, S.~K. Venkataramanaiah, J.~sun Seo, and
  M.~Mattina, ``{FixyNN: Efficient Hardware for Mobile Computer Vision via
  Transfer Learning},'' in \emph{Proceedings of the 2nd SysML Conference, Palo
  Alto, CA, USA}, 2019.

\bibitem{interstellar-asplos20}
\BIBentryALTinterwordspacing
X.~Yang, M.~Gao, Q.~Liu, J.~Setter, J.~Pu, A.~Nayak, S.~Bell, K.~Cao, H.~Ha,
  P.~Raina, C.~Kozyrakis, and M.~Horowitz, ``Interstellar: Using halide's
  scheduling language to analyze dnn accelerators,'' in \emph{Proceedings of
  the Twenty-Fifth International Conference on Architectural Support for
  Programming Languages and Operating Systems}, ser. ASPLOS '20.\hskip 1em plus
  0.5em minus 0.4em\relax New York, NY, USA: Association for Computing
  Machinery, 2020, p. 369–383. [Online]. Available:
  \url{https://doi.org/10.1145/3373376.3378514}
\BIBentrySTDinterwordspacing

\bibitem{zhang2020sparch}
Z.~Zhang, H.~Wang, S.~Han, and W.~J. Dally, ``Sparch: Efficient architecture
  for sparse matrix multiplication,'' in \emph{26th IEEE International
  Symposium on High Performance Computer Architecture (HPCA)}, 2020.

\bibitem{zhou-micro18}
\BIBentryALTinterwordspacing
X.~Zhou, Z.~Du, Q.~Guo, S.~Liu, C.~Liu, C.~Wang, X.~Zhou, L.~Li, T.~Chen, and
  Y.~Chen, ``Cambricon-s: Addressing irregularity in sparse neural networks
  through a cooperative software/hardware approach,'' in \emph{Proceedings of
  the 51st Annual IEEE/ACM International Symposium on Microarchitecture}, ser.
  MICRO-51.\hskip 1em plus 0.5em minus 0.4em\relax IEEE Press, 2018, p.
  15–28. [Online]. Available: \url{https://doi.org/10.1109/MICRO.2018.00011}
\BIBentrySTDinterwordspacing

\bibitem{zhu2017prune}
\BIBentryALTinterwordspacing
M.~Zhu and S.~Gupta, ``To prune, or not to prune: exploring the efficacy of
  pruning for model compression.'' \emph{CoRR}, vol. abs/1710.01878, 2017.
  [Online]. Available:
  \url{http://dblp.uni-trier.de/db/journals/corr/corr1710.html#abs-1710-01878}
\BIBentrySTDinterwordspacing

\bibitem{euphrates}
\BIBentryALTinterwordspacing
Y.~Zhu, A.~Samajdar, M.~Mattina, and P.~Whatmough, ``{Euphrates: Algorithm-SoC
  Co-Design for Low-Power Mobile Continuous Vision},'' in \emph{Proceedings of
  the 45th Annual International Symposium on Computer Architecture}, ser. ISCA
  '18.\hskip 1em plus 0.5em minus 0.4em\relax IEEE Press, 2018, p. 547–560.
  [Online]. Available: \url{https://doi.org/10.1109/ISCA.2018.00052}
\BIBentrySTDinterwordspacing

\end{thebibliography}
